\documentclass[conference]{IEEEtran}
\IEEEoverridecommandlockouts
\usepackage{cite}
\usepackage{amsmath,amssymb,amsfonts}
\usepackage{algorithmic}
\usepackage{graphicx}
\usepackage{textcomp}
\usepackage[dvipsnames]{xcolor}

\usepackage{tabu}
\usepackage{booktabs}    
\usepackage{array}
\usepackage{tabulary}
\usepackage{hyperref}
\usepackage{subfigure}

\def\BibTeX{{\rm B\kern-.05em{\sc i\kern-.025em b}\kern-.08em
    T\kern-.1667em\lower.7ex\hbox{E}\kern-.125emX}}
\begin{document}
\newcommand{\alex}[1]{{\color{cyan}{\textbf #1}}}
\title{Visualization Resources: A Starting Point\\
}


\author{\IEEEauthorblockN{Xiaoxiao Liu\textsuperscript{1}
Mohammad Alharbi\textsuperscript{2} Joe Best\textsuperscript{3} Jian Chen\textsuperscript{4} Alexandra Diehl\textsuperscript{5} Elif Firat\textsuperscript{3} \\ Dylan Rees\textsuperscript{6} Qiru Wang\textsuperscript{3} Robert S Laramee\textsuperscript{3} }\\

\IEEEauthorblockA{\textsuperscript{1}\textit{Bournemouth University}, UK, 
XLiu@bournemouth.ac.uk\\
\textsuperscript{2}\textit{Swansea University}, UK, 
508205@swansea.ac.uk\\
\textsuperscript{3}\textit{University of Nottingham}, UK, 
\{Joseph.Best, Elif.Firat, Qiru.Wang, Robert.Laramee\}@nottingham.ac.uk\\
\textsuperscript{4}\textit{The Ohio State University}, US, 
Chen.8028@osu.edu\\
\textsuperscript{5}\textit{University of Zurich}, Switzerland, 
Diehl@ifi.uzh.ch \\
\textsuperscript{6}\textit{Zuken UK Ltd.}, UK, 
849119@swansea.ac.uk} 
}

\maketitle

\begin{abstract}
    Visualization, as a vibrant field for researchers, practitioners, and higher educational institutions, is growing and evolving very rapidly.  Tremendous progress has been made since 1987, the year often cited as the beginning of data visualization as a distinct field.  As such, the number of visualization resources and the demand for those resources are increasing at a very fast pace. We present a collection of open visualization resources for all those with an interest in interactive data visualization and visual analytics. Because the number of resources is so large, we focus on collections of resources, of which there are already very many ranging from literature collections to collections of practitioner resources. We develop a novel classification of visualization resource collections based on the resource type, e.g. literature-based, web-based, etc. The result is a helpful overview and details-on-demand of many useful resources.  The collection offers a valuable jump-start for those seeking out data visualization resources from all backgrounds spanning from beginners such as students to teachers, practitioners, and researchers wishing to create their own advanced or novel visual designs.
\end{abstract}


\section{Introduction and Motivation}
Data visualization has become an increasingly important solution for analyzing and exploring huge volumes of complex data. As such, the number of data visualization resources has grown very rapidly over the past decades. However, for students, researchers, practitioners, or visualization scientists who work in the field, searching for data visualization resources can be challenging and time consuming due to the rapidly evolving landscape of visualization
and because a simple Google search will not always lead to the most valuable resources.

This collection of visualization resources serves as an effective starting point when searching for both literature and tools related to visualization due to the many hours of searching and curating dedicated to this project and decades of collective teaching and research experience.
Many students, teachers, practitioners, institutions and companies have been collecting data visualization resources, such as Souto \cite{FabioSouto} who provides a curated list of valuable open-source data visualization developer resources such as programming libraries, software and websites. These collections focus on the tools and techniques of visualization. Survey papers and literature, such as Lam et al.~\cite{Lam2012a} and Isenberg et al.~\cite{isenberg2013systematic} provide a valuable collection of meta-data on research papers at the annual conference series related to evaluating visualization.

This paper aims to provide a novel overview of collections of data visualization resources. It also provides a categorization of those resources. The contributions of this unique survey include:
\begin{itemize}
    \item The first survey of its kind on collections of resources for data visualization,
    \item A novel categorization of visualization resource collections organized around readership with a focus on a students, researchers, and practitioners.
\end{itemize}
Our web-based collection of visualization resources is available at: \href{https://sites.google.com/view/visres/}{https://sites.google.com/view/visres/}.
\paragraph*{\textbf{Survey Scope}}
By \textit{resource} we mean content that provides benefits to visualization students, practitioners, and researchers.
The inspiration behind this project comes from real-life
experience.
We are often asked what visualization resources are available from students, researchers, and practitioners.
This is a response to that common question.

We focus on free, visualization-specific resources, e.g.,
open collections of visualization images or other collective meta-data.

We catalog collections of visualization resources  (as opposed  to  individual  items). e.g. A survey of surveys  and  state-of-the-art  reports rather than an  individual survey, or a survey of books rather than an individual book.
There are too many visualization research papers (thousands) to list individually. However, there are resources that present and explore the large collection of visualization literature such as VisPubData \cite{Isenberg2017a}, VisImageNavigator~\cite{chen2020vis30k}, and CiteVis \cite{Stasko2013}.  Similarly, there are over 100 survey papers on the topic of visualization and visual analytics.  Therefore, we focus on resources such as surveys of surveys\cite{Alharbi2018}\cite{Alharbi2017}\cite{McNabb2017c}. We present collections of visualization resources spanning different types such as literature and websites that gather resources, together as a collection. In summary, we include collections of visualization resources we believe provide great value to students, researchers, and practitioners.

We prioritize collections of free open-source visualization resources that benefit visualization students, practitioners, and researchers. We focus on collections offered by non-profit institutions and organizations. This includes higher educational institutions, non-profit institutions such as Wikipedia, and collections of visualization resources gathered together for public use by volunteers.

Finally, in order to make the scope manageable, we ensure that the resources have a focus on visualization.

\paragraph*{\textbf{The Search Process}}

The search process for visualization resources is challenging, time consuming, and cannot be solved by a simple Google search.
Although, we do incorporate Google search for some collections of visualization resources.
Our collection process consists of several aspects, including teaching a visualization course where students themselves search for resources, experience in reading and reviewing visualization research papers, attending visualization events such as conferences and workshops, and of course traditional search engine queries. This kind of process is difficult to systematize and is the result of a team of visualization researchers and students with more than 20 collective years of search labor.
The fruits of years worth of search labor benefits the reader. 
We provide more details on how we search for resources in each sub-section of the paper.

\paragraph*{\textbf{The Benefits of Collections}}

Focusing on collections offers a number of benefits. Firstly, a collection of collections serves as an effective approach to navigate the abundance of visualization resources. Secondly, resource collections are often developed and maintained by an individual or a team of curators.
Thirdly, by focusing on collections, we take advantage of the many hours of search and construct labor already invested in collecting the resources.
Fourthly, by focusing on collections, we inherit the benefits that bring similar resources together and their respective categorization.


\paragraph*{\textbf{Categorization:}}
There are many different ways to group  and categorize visualization resources. One possibility is to classify them based on type. For example, collections of refereed research papers with accompanying online resources, visualization books with online resources, open-source collections of visualization software, non-profit websites with collections of visualization resources, etc. Another possibility is to categorize the resource collections based on visualization sub-field. For example, visual analytics, information visualization, or scientific visualization. However, as we shall see, it would be difficult to categorize resources this way due to subject cross-over.
A third categorization could be based on target audience such as resource collections for students, researchers, practitioners, or visualization scientists. However, many resource collections are valuable to multiple types of users. Another possibility is to classify resource collections based on special subjects such as education, geospatial visualization, or data-centered resources. However, this might result in many different categories. 

After considering several different categorization schemes,
we have chosen a grouping based on resource type because we believe this is the most relevant with respect to the prospective readers' interest.
In other words, resource type often aligns with the different categories of readership.
For example, a researcher may be interested in finding starting points for relevant refereed literature.
Thus, we group the collections of related refereed literature together.
In another example, a student may specifically be interested in data visualization web sites that offer a helpful collection of resources.
Therefore, we have grouped the related web sites together.
Attempting to categorize visualization resources by subject will result in large crossover between categories.
Ultimately no categorization is perfect in this case, i.e. results in no crossover.

\paragraph*{\textbf{Related Work}}

The web site usabiliTEST \cite{UsabiliTEST} is an online tool for usability testing and information architecture on web usability issues. The website provides a page called, “Methods table” which provides a collection of resources on usability. It is a very valuable collection of resources for the Human-Computer Interaction (HCI) community. This is the closest related work we found on collections of resources. To our knowledge, no refereed paper has been published on this topic.

\section{Resources Focused on Collections of Refereed Literature}

This category focuses on peer-reviewed literature that offers collections of research papers.
It offers a helpful starting point for readers interested in obtaining starting points for literature overviews.
In this section, resources on surveys of surveys are presented. Next, research papers with online collections and resources are proposed. In the third sub-section, we provide some surveys which gather online collections of images. The fourth sub-section focuses on SurVis resources discussed in the literature. Lastly, surveys of visualization books are discussed.  


\paragraph*{\textbf{Refereed Survey-based Resources (Surveys of Surveys)}}

Surveys present a valuable means to quickly find previous research on a particular topic, however, there are a growing number of topics and therefore a growing number of surveys. To address this, there have been recent developments on surveys of surveys\cite{Alharbi2019a} \cite{Alharbi2017} \cite{McNabb2017c} \cite{Min}. These surveys of surveys are valuable resources offering a concise overview of the interactive data visualization and visual analytic fields.

\textbf{The Search Process:}
In order to find these surveys, we searched for the phrase ``visualization survey of surveys" and a number of variants on this phrase.  After our initial findings we then did two further searches.
The first one is a recursive search, i.e., we examined the references
found in the initial survey of surveys.
The second search is a forward looking search, i.e., we utilized Google's ``cited by" feature for each survey of surveys.
For the rest of this paper we refer to these searches as the recursive and forward-looking searches.
\textbf{Inclusion Criteria:}
We include each survey of surveys we find on the topic of visualization or visual analytics. 

This category was pioneered with the work of McNabb and Laramee who provide a comprehensive survey of 85 information visualization and visual analytics surveys\cite{McNabb2017c}. This work is accompanied by a SurVis based web resource, listing a large collection of surveys and state-of-the-art reports.

Text visualisation has seen rapid gain in popularity in recent years, accompanied by a number of survey papers on the topic. Alharbi and Laramee provide an overview of these with a review of 14 text-based surveys\cite{Alharbi2019a}. Similarly, the surveys from the cross-disciplinary field of molecular dynamics have been analyzed by Alharbi et al.\cite{Alharbi2017}.

A premier journal for publishing surveys in data visualization and computer graphics, Computer Graphics Forum, has an extensive collection of surveys -- every survey published in Computer Graphics Forum. The website by Chen \cite{Min} features, over 180 survey papers presented in chronological order dating back to 1985.

As a hot topic in comprehending patterns and predicting trends of data, machine learning models have developed fast in many different areas. Accordingly, information visualization of machine learning models provides effective solution in interpreting the workings of these models. Inspired by the work of McNabb and Laramee\cite{McNabb2017c}, Chatzimparmpas et al.\cite{Chatzimparmpas2020} contributes a survey of surveys on the exploration and interpretability of machine learning models. This work contributes 18 papers related to the visualization of machine learning (2014 - 2018).


\paragraph*{\textbf{Research Papers with Online Meta-data Collections and Resources}}

There are a number of previous related papers that examine collective meta-data from published visualization papers and publish the meta-data itself as a valuable contribution for further research and analysis.

\textbf{The Search Process:}
In order to find refereed literature offering a collection of resources, we started with research papers that we read first-hand offering meta-data collections.
We came across these research papers through regular reading and reviewing of research literature, consistently attending visualization conferences and related events, and personal communication with other researchers in the field.
Moreover, these findings are based on a collective experience of many years.
In other words, the search is based on a team effort by all the co-authors of this manuscript.
After our initial findings based on reading and experience, we then performed two further searches for each research paper in this category: the recursive search and the forward looking search described previously.
\textbf{Inclusion Criteria:}
We include each visualization paper that offers a valuable collection of meta-data that facilitates both search and comparison of visualization literature on a collective level. 

Lam et al. analyze information visualization papers and classify their evaluation types into seven categories. They published the full list of information visualization papers along with their classifications online \cite{Lam2012a}. This paper was later extended to cover scientific visualization papers~\cite{isenberg2013systematic}.
Stakso et al. published an online tool called Citevis that shows which visualization papers are cited by others using interaction \cite{Stasko2013}. Isenberg et al. examine and classify the types of contribution made by each visualization paper and publish the full classification online\cite{Isenberga}. Isenberg et al. also examine and classify the keywords used in each visualization research paper and study how they compare to keywords used by a typical visualization research paper submission system \cite{Isenberg2017b}. They accompany their analysis with an online keyword browser.

Conceptually, the work we present here extends that of Isenberg et al.\cite{Isenberg2017a} who collected and curated a dataset featuring every visualization conference paper ever published. The collection is available online at VisPubData.org.

Matejka and Fitzmaurice implement a method to create a dataset slightly modifying an already existing one \cite{Justin}. While preserving their e.g. mean, standard, deviation etc., the statistical properties maintained are the same. Their work was based on the theory of Anscombe’s Quartet presented by F.J. Anscombe \cite{Anscombe1973} and the Datasaurus Dozen created by Alberto Cairo \cite{AlbertoCairo}. Based on these methods, a web page is also provided to demonstrate the variation of datasets \cite{Matejka2017}.

Information Visualization Evaluation Using Crowdsourcing, by Borgo et al., \cite{Borgo} contains a spreadsheet of 82 papers and the crowd-sourcing meta-data ralated to each paper.

\paragraph*{\textbf{Survey Papers with Accompanying Image Collection Browsers}}\label{sec:image}

\begin{table*}[!tb]
    \caption{Image Browser References: A summary of the image collection browsers described in Section~\ref{sec:image}.  These papers feature a quality collection of visualization-related images.  See Figure~\ref{fig:image} for images of the image collection web pages}
    \centering

    \resizebox{\textwidth}{!}{
        \begin{tabulary}{\textwidth}{|l|l|}
            \hline

            \textbf{Image Browser Name}  &  \textbf{URL}  \\
            \hline

            \textbf{Treevis.net} \cite{Schulz2011} & \url{https://treevis.net/}   \\
            \hline

            \textbf{TimeViz Browser}\cite{Aigner2011a} & \url{https://vcg.informatik.uni-rostock.de/~ct/timeviz/timeviz.html} \\
            \hline
            \textbf{Multiviz.net} \cite{Kehrer2013a} & \url{https://multivis.net/} \\

            \hline
            \textbf{Text Visualization Browser} \cite{Kucher2015} & \url{https://textvis.lnu.se/}   \\
            \hline

            \textbf{BioVis Explorer} \cite{Kerren2017a} & \url{https://biovis.lnu.se/}   \\ \hline

            \textbf{SentimentVis Browser} \cite{Kucher2018a} & \url{https://sentimentvis.lnu.se/}   \\
            \hline

            \textbf{TrustMLVis Browser} \cite{Chatzimparmpas2020} & \url{https://trustmlvis.lnu.se/}   \\
            \hline

            \textbf{Finance Vis Browser} \cite{Dumas2014} & \url{http://financevis.net/}   \\
            \hline

            \textbf{Predictive Visual Analytics Browser} \cite{Lu2017a} & \url{http://104.196.253.120/pva_browser/}   \\
            \hline

            \textbf{MVN Visualization Techniques} \cite{Nobre2019b} & \url{https://vdl.sci.utah.edu/mvnv/}   \\
            \hline

            \textbf{UncertaintyViz Browser} \cite{Jena2020} & \url{https://amitjenaiitbm.github.io/uncertaintyVizBrowser/}   \\
            \hline

            \textbf{VisImageNavigator} \cite{chen2020vis30k} & \url{https://visimagenavigator.github.io/}   \\
            \hline

        \end{tabulary}
    }
    \label{tab:image}
\end{table*}

\begin{figure*}[!tb]
    \centering
    \subfigure{
        \begin{minipage}[t]{0.237\linewidth}
            \centering
            \includegraphics[width=1.7in]{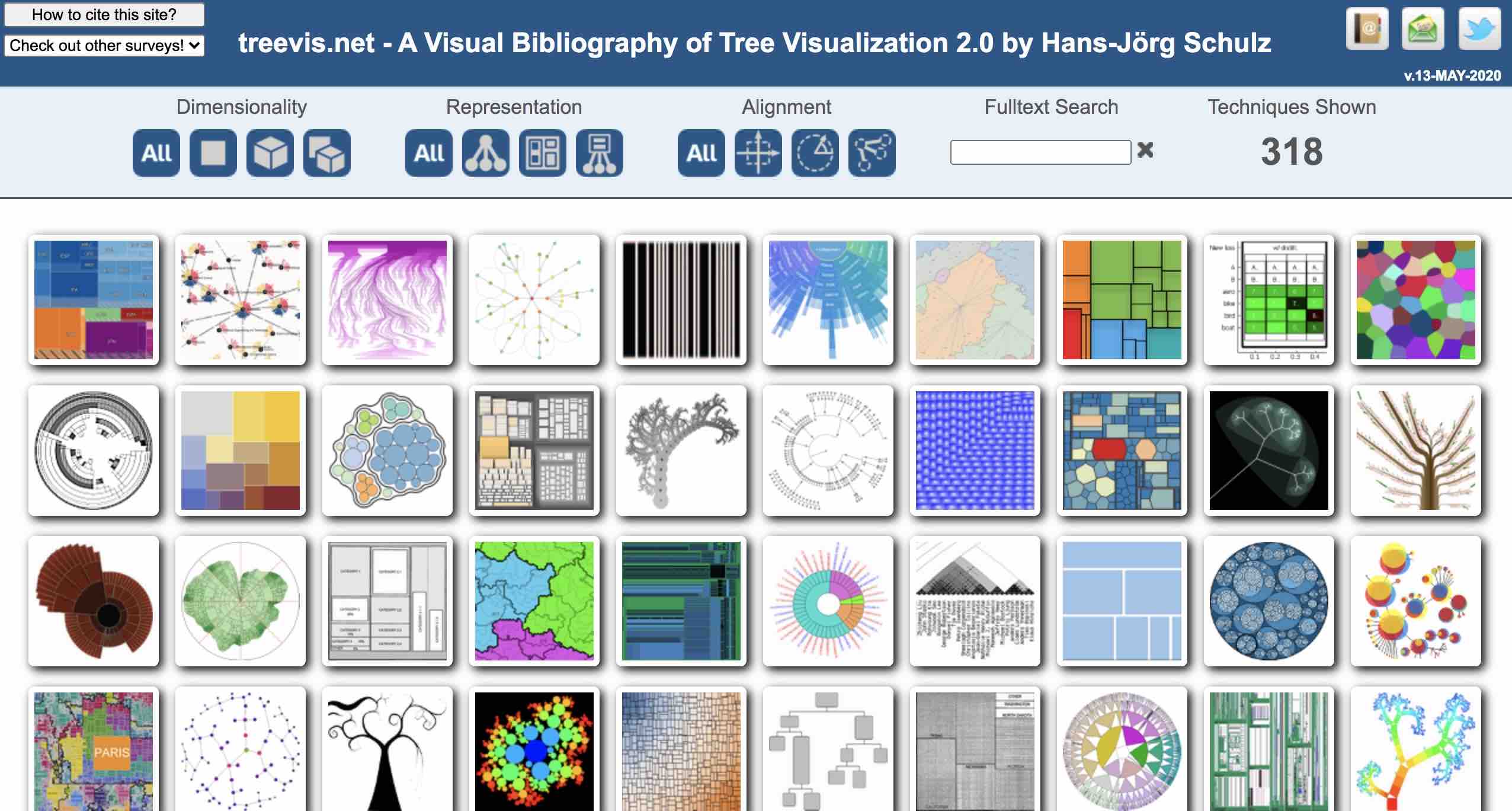}
        \end{minipage}%
        \begin{minipage}[t]{0.237\linewidth}
            \centering
            \includegraphics[width=1.7in]{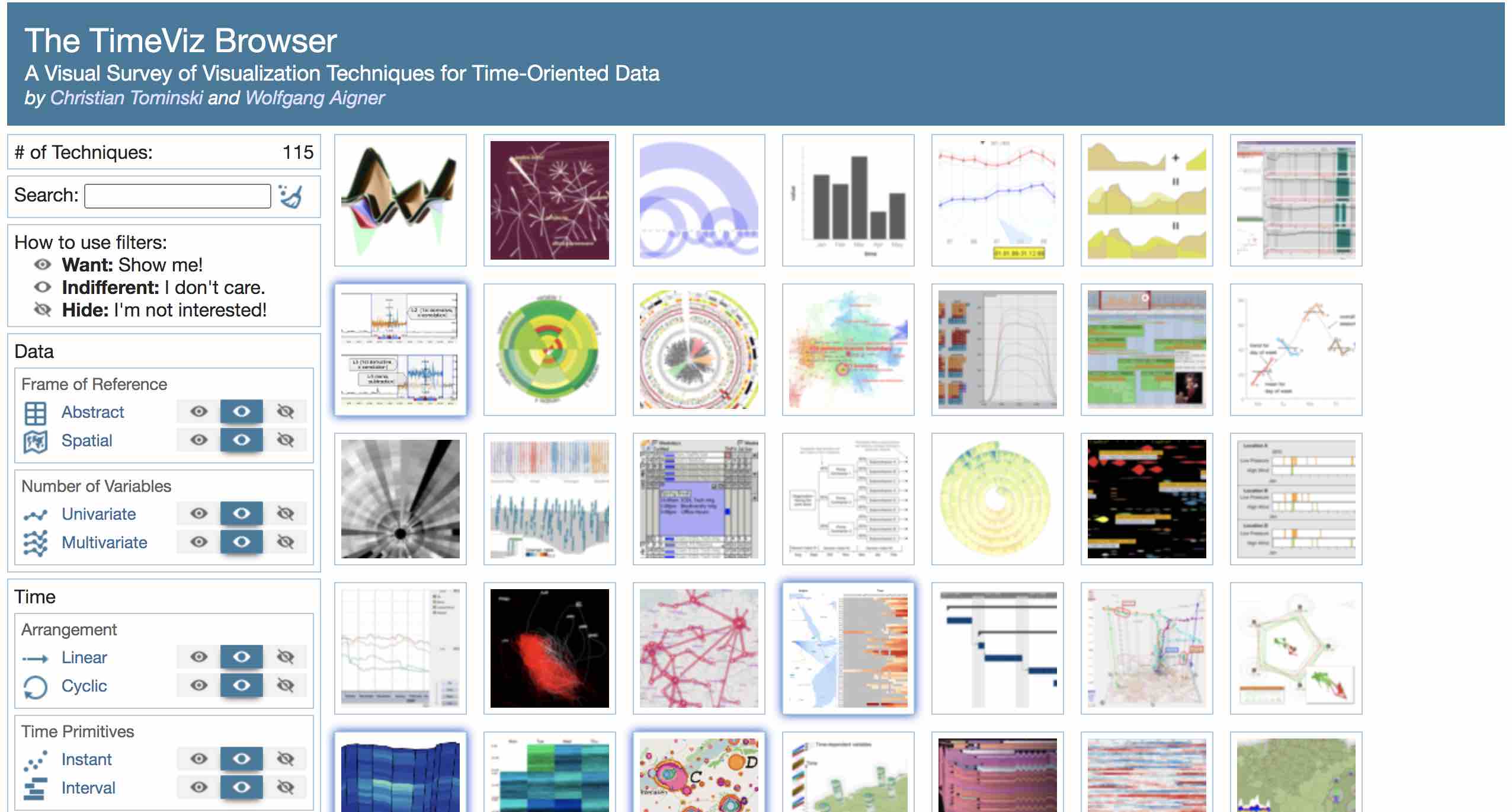}
        \end{minipage}%
        \begin{minipage}[t]{0.237\linewidth}
            \centering
            \includegraphics[width=1.7in]{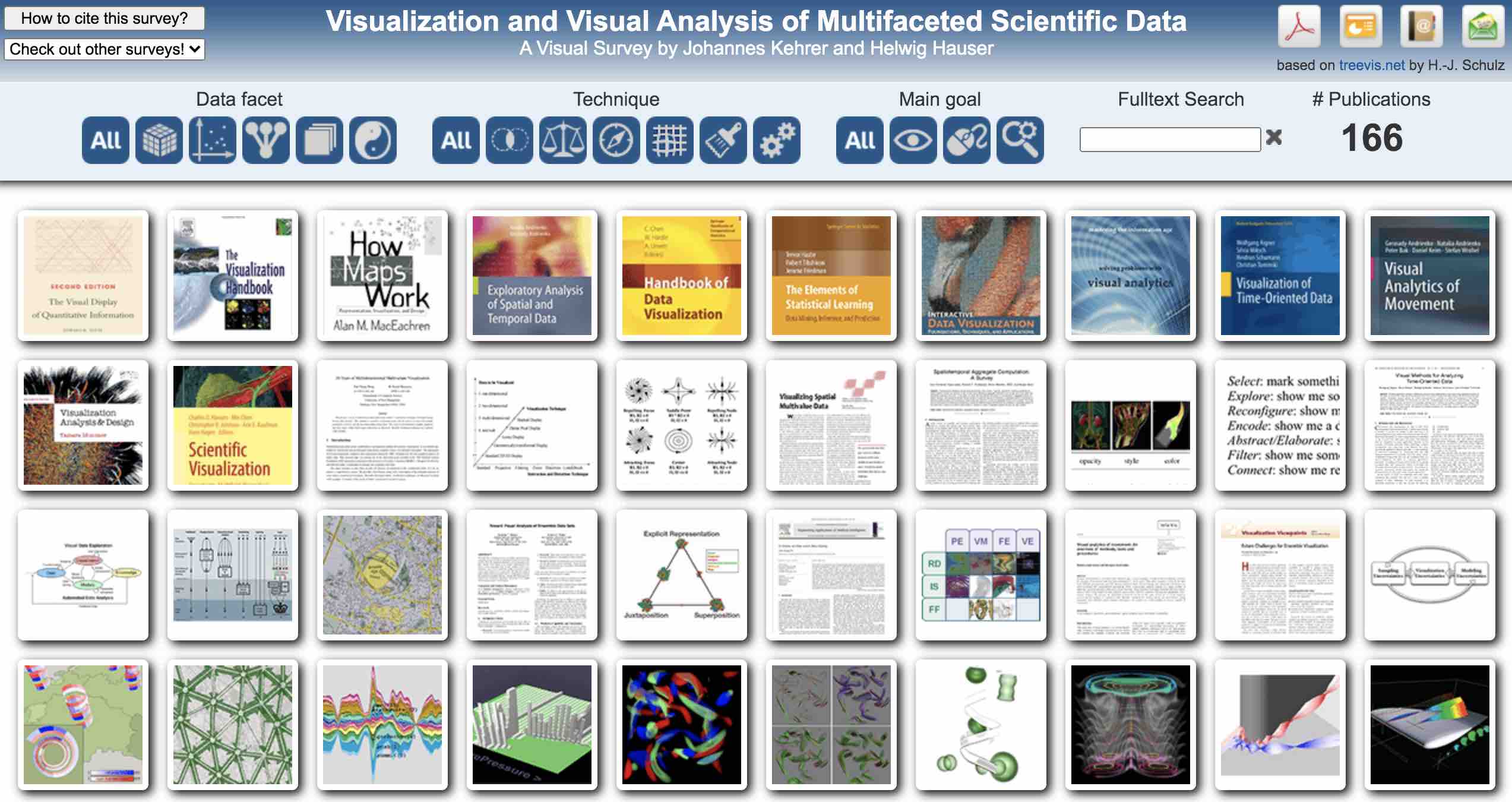}
        \end{minipage}%
        \begin{minipage}[t]{0.237\linewidth}
            \centering
            \includegraphics[width=1.7in]{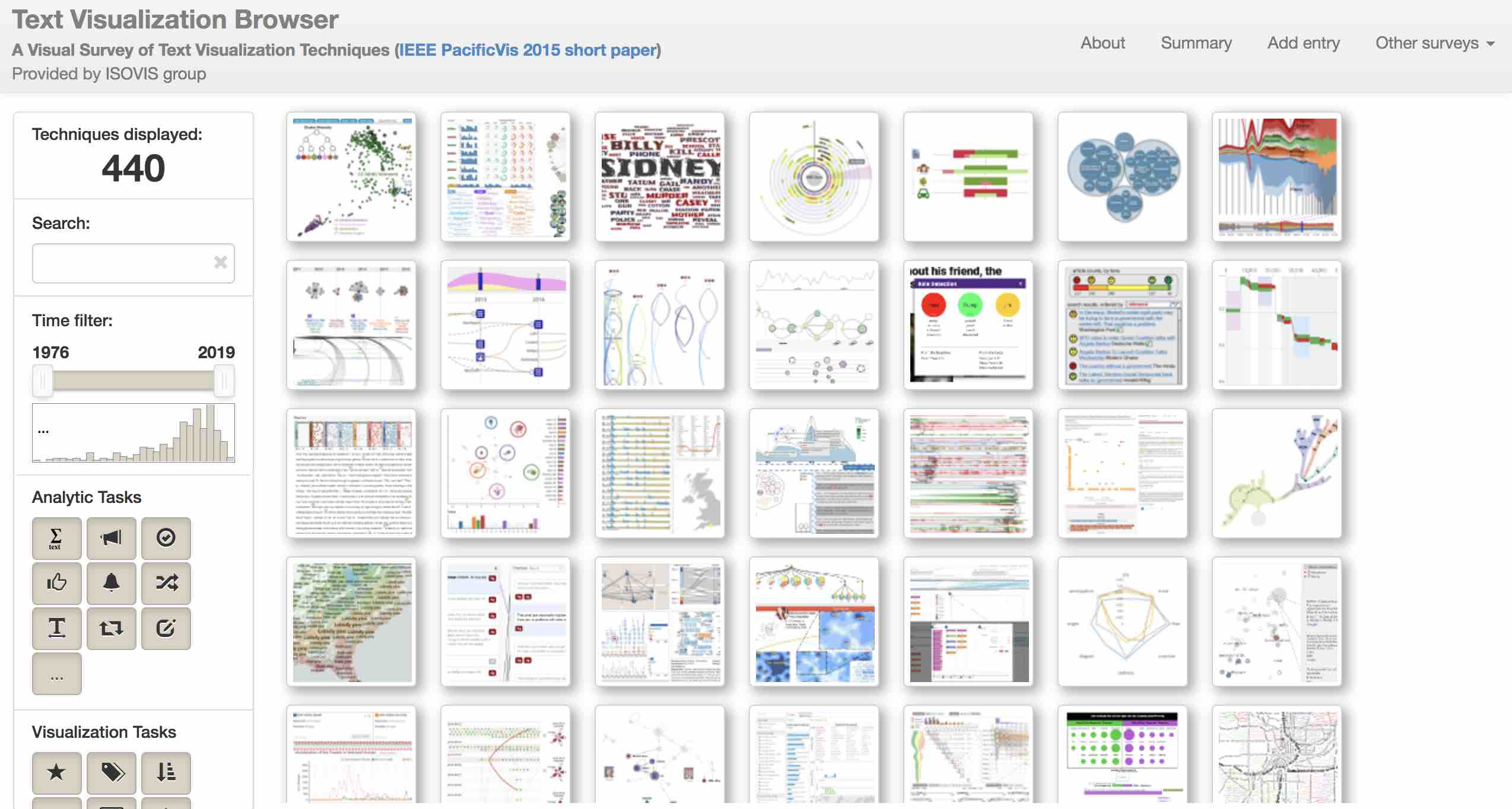}
        \end{minipage}%
    }%

    \subfigure{
        \begin{minipage}[t]{0.237\linewidth}
            \centering
            \includegraphics[width=1.7in]{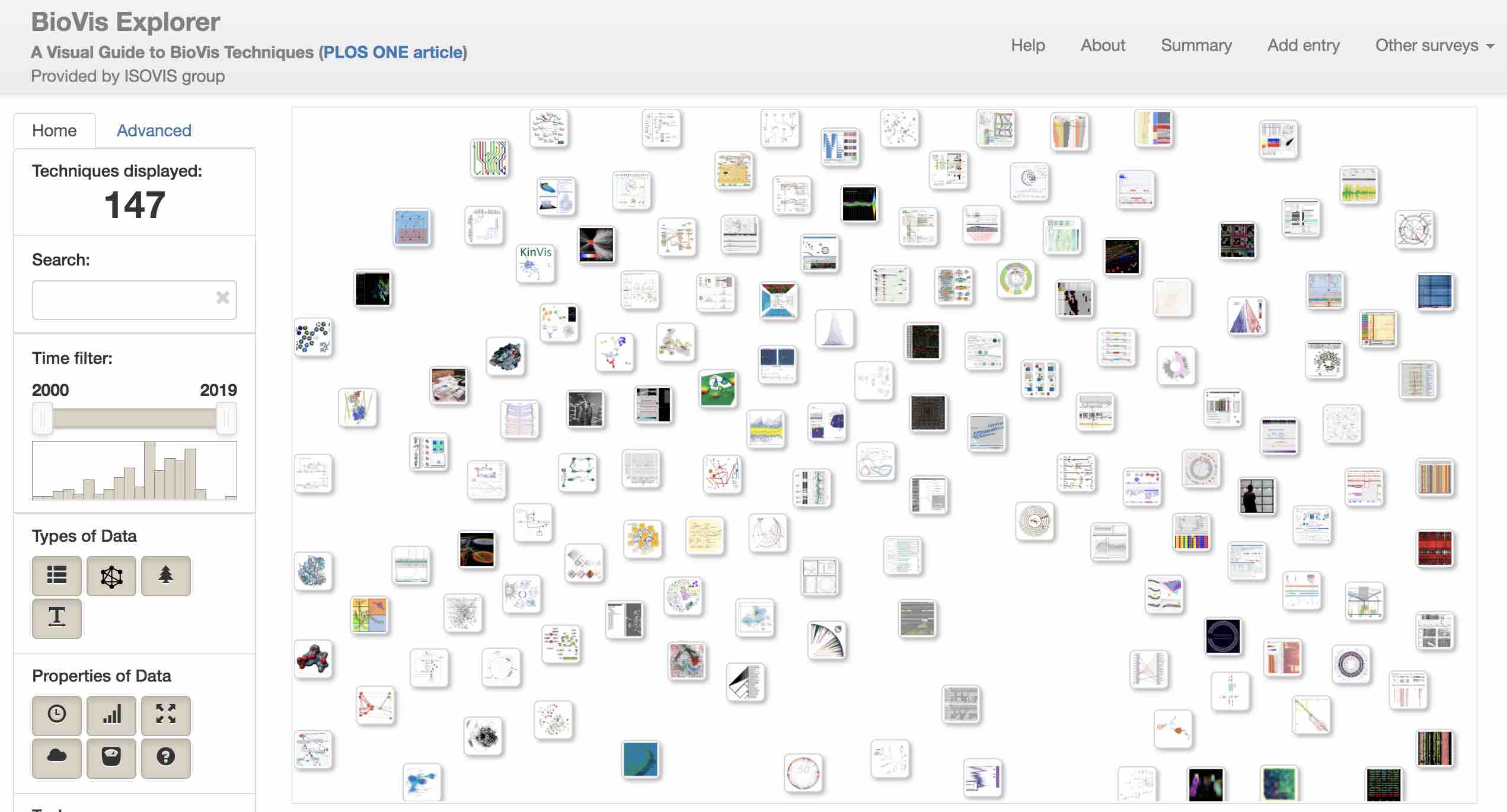}
        \end{minipage}%
        \begin{minipage}[t]{0.237\linewidth}
            \centering
            \includegraphics[width=1.7in]{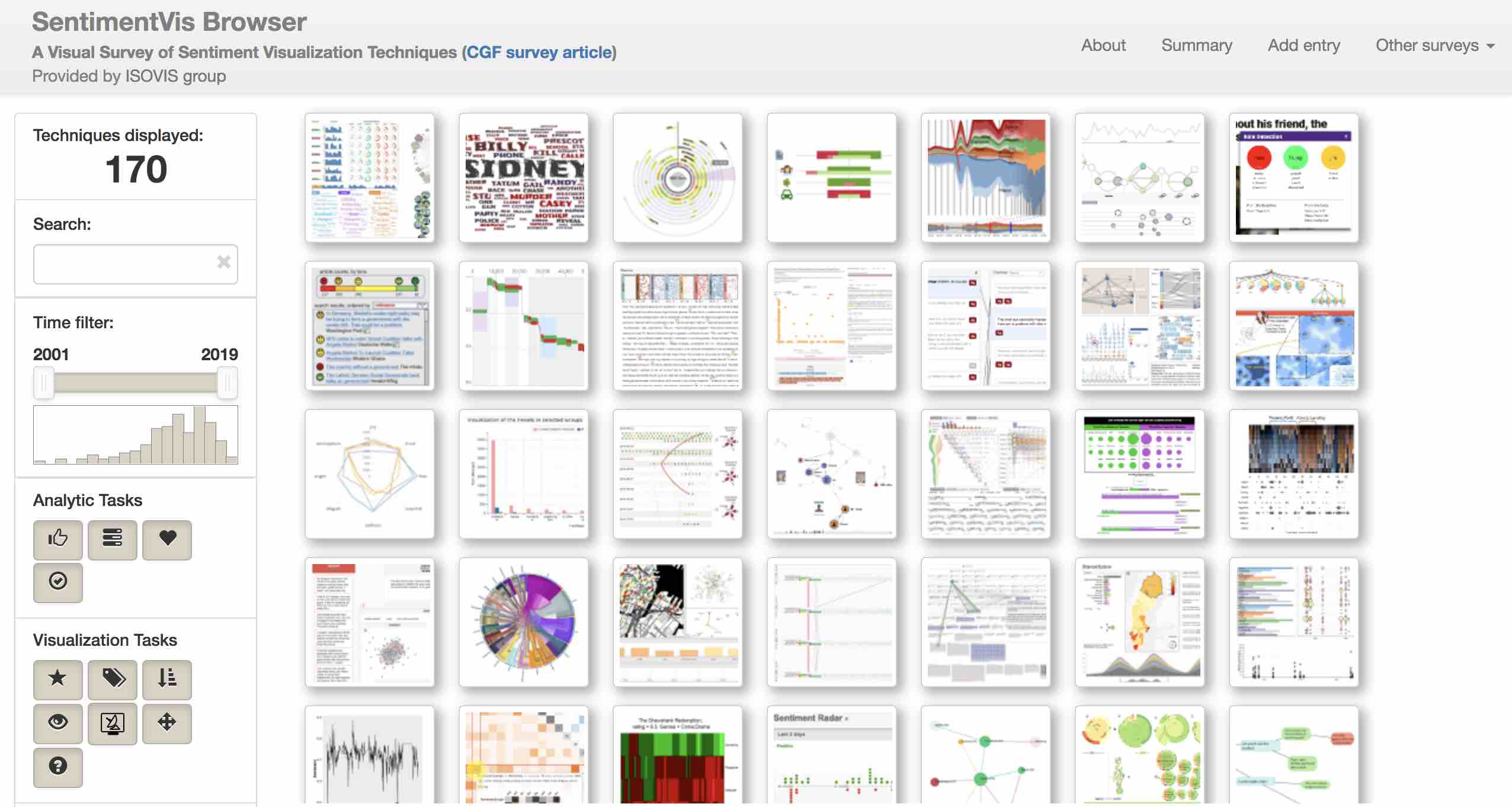}
        \end{minipage}%
        \begin{minipage}[t]{0.237\linewidth}
            \centering
            \includegraphics[width=1.7in]{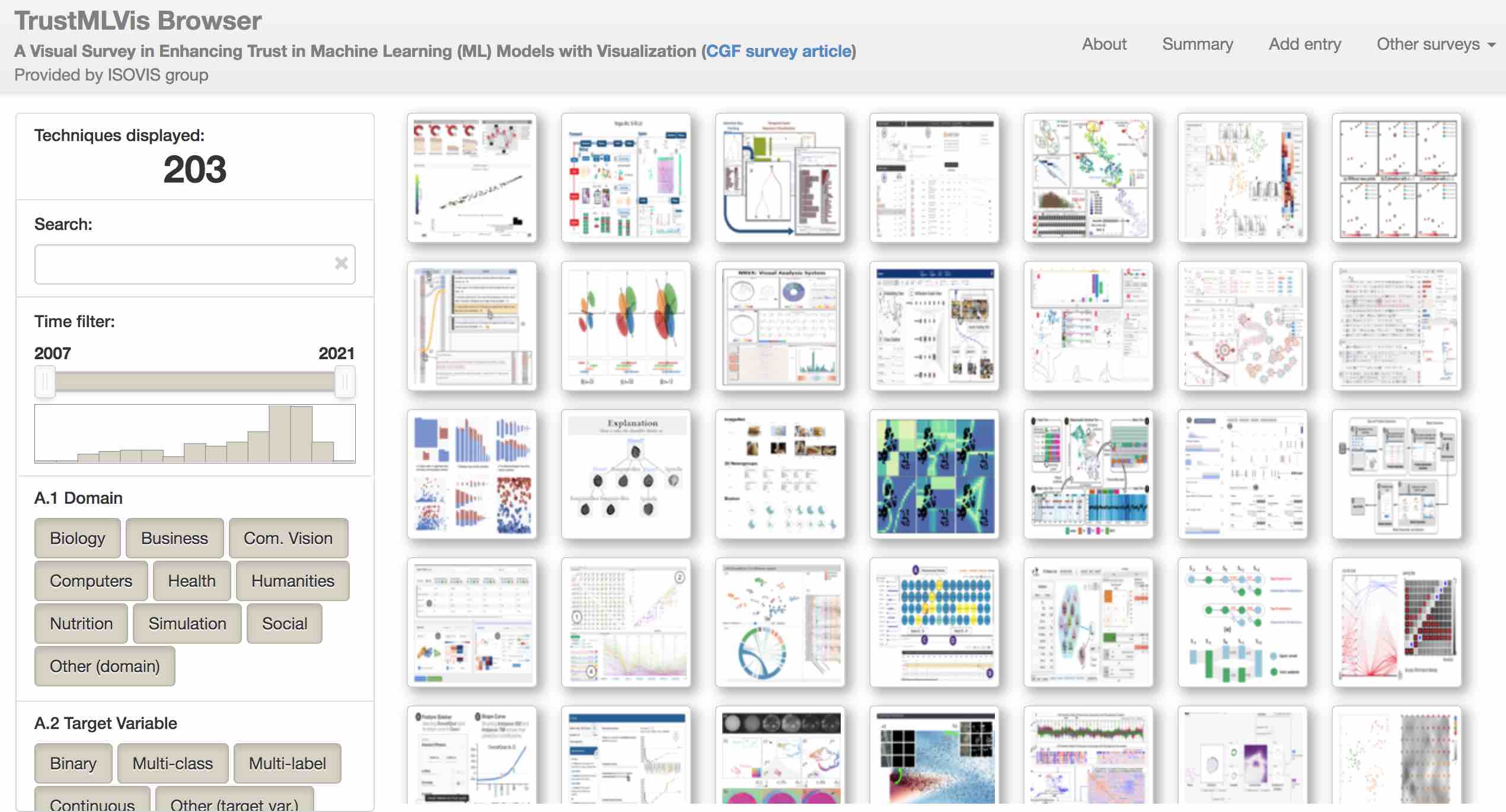}
        \end{minipage}%
        \begin{minipage}[t]{0.237\linewidth}
            \centering
            \includegraphics[width=1.7in]{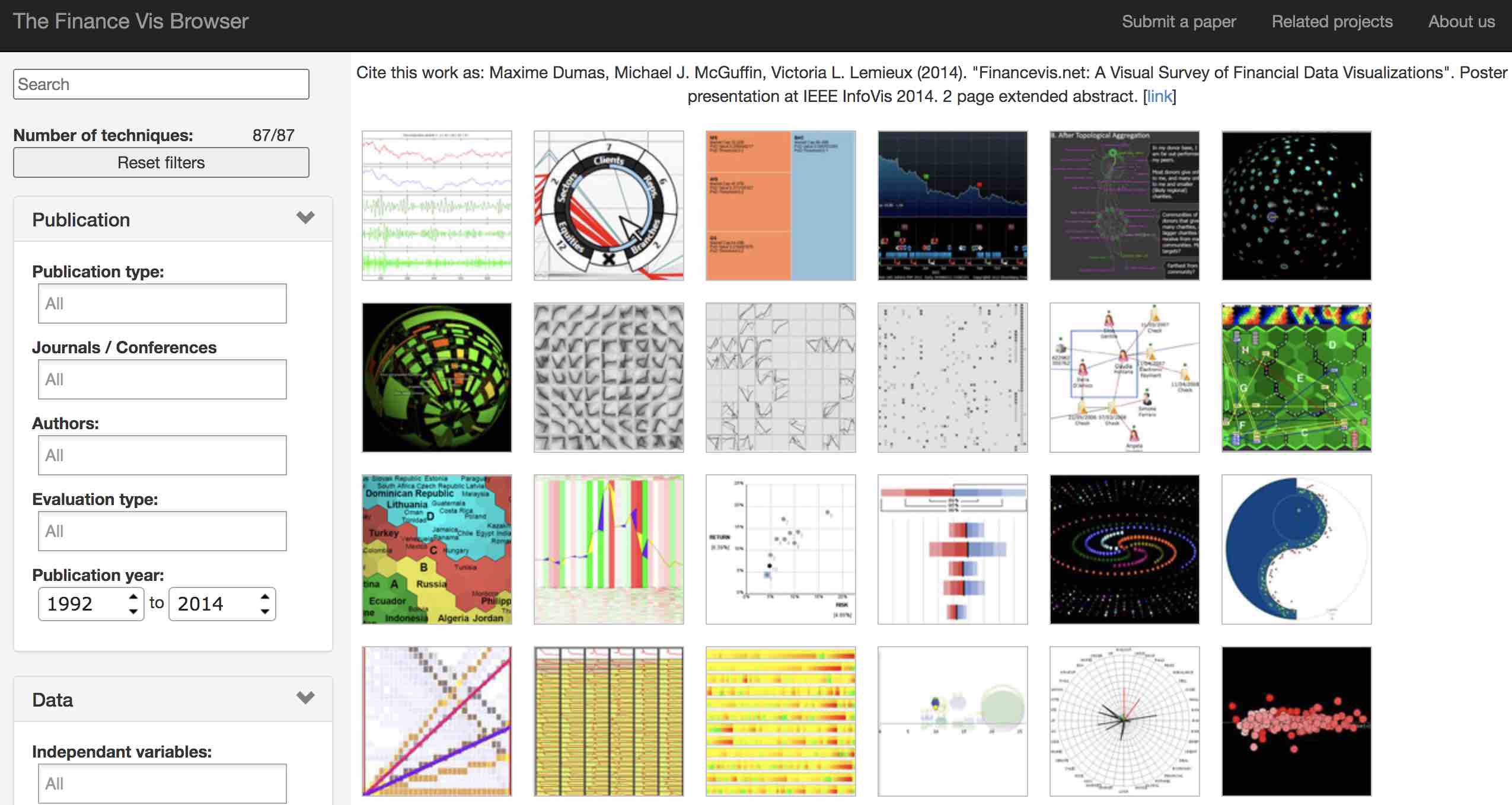}
        \end{minipage}%
    }%

    \subfigure{
        \begin{minipage}[t]{0.237\linewidth}
            \centering
            \includegraphics[width=1.7in]{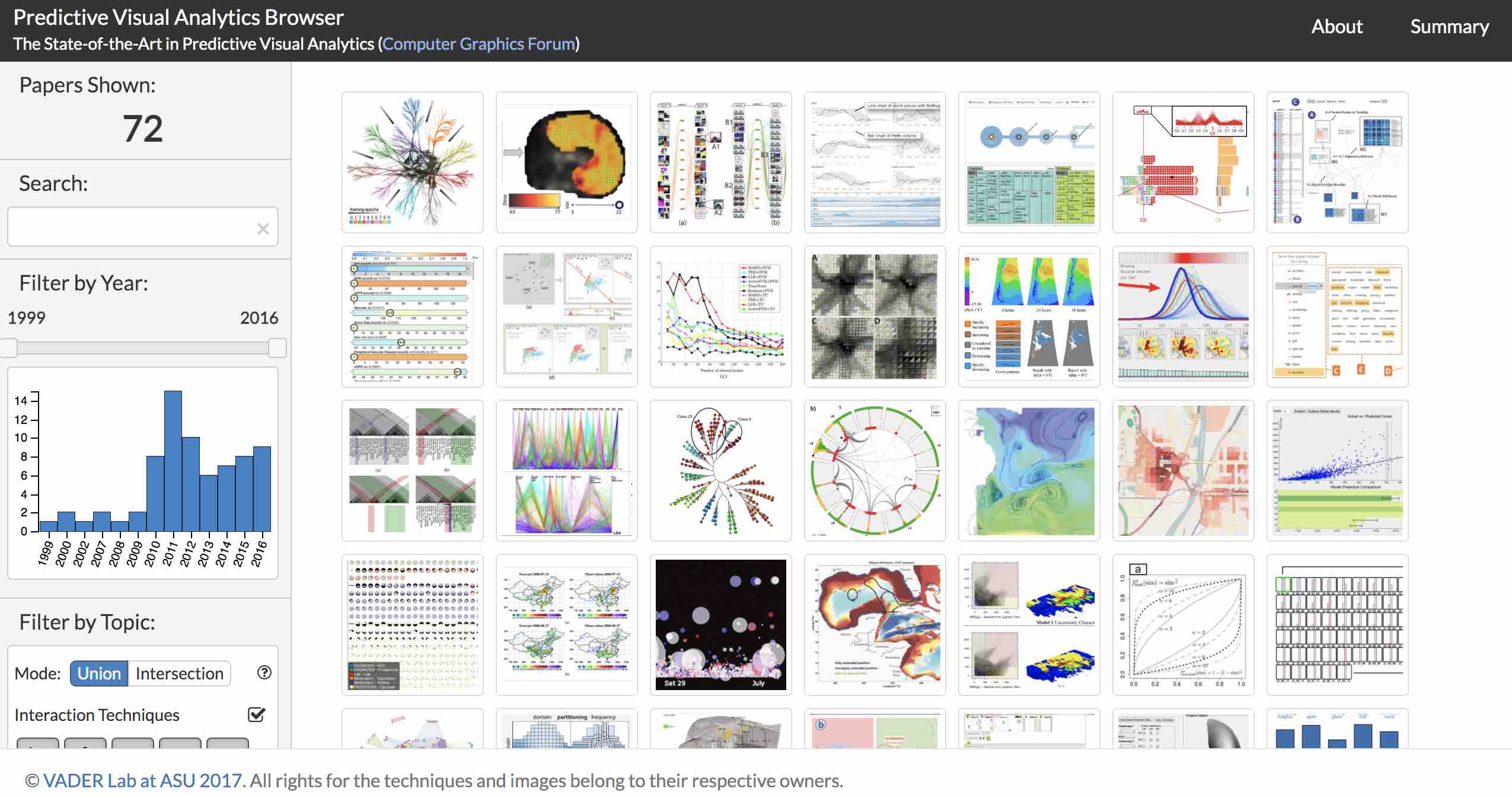}
        \end{minipage}%
        \begin{minipage}[t]{0.237\linewidth}
            \centering
            \includegraphics[width=1.7in]{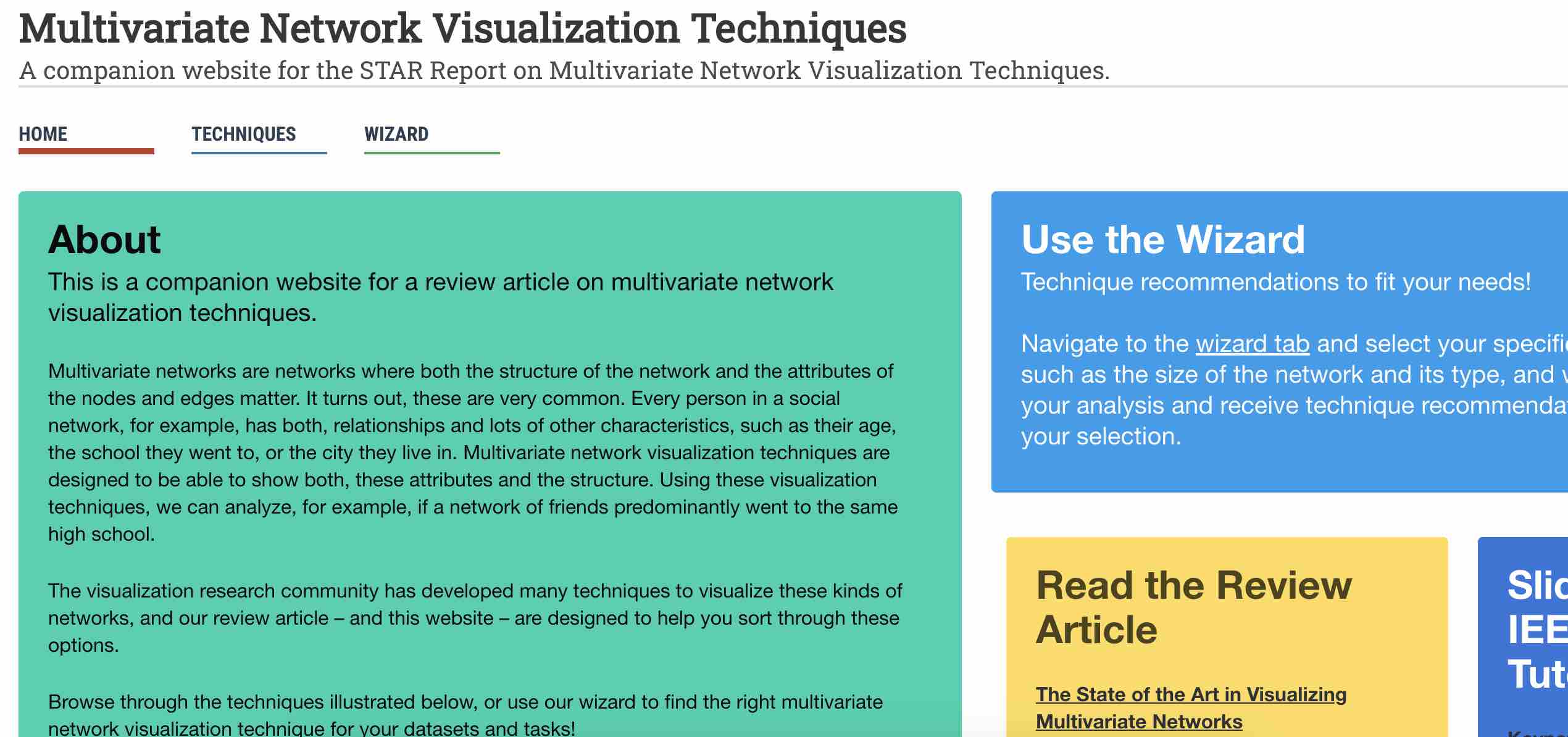}
        \end{minipage}%
        \begin{minipage}[t]{0.237\linewidth}
            \centering
            \includegraphics[width=1.7in]{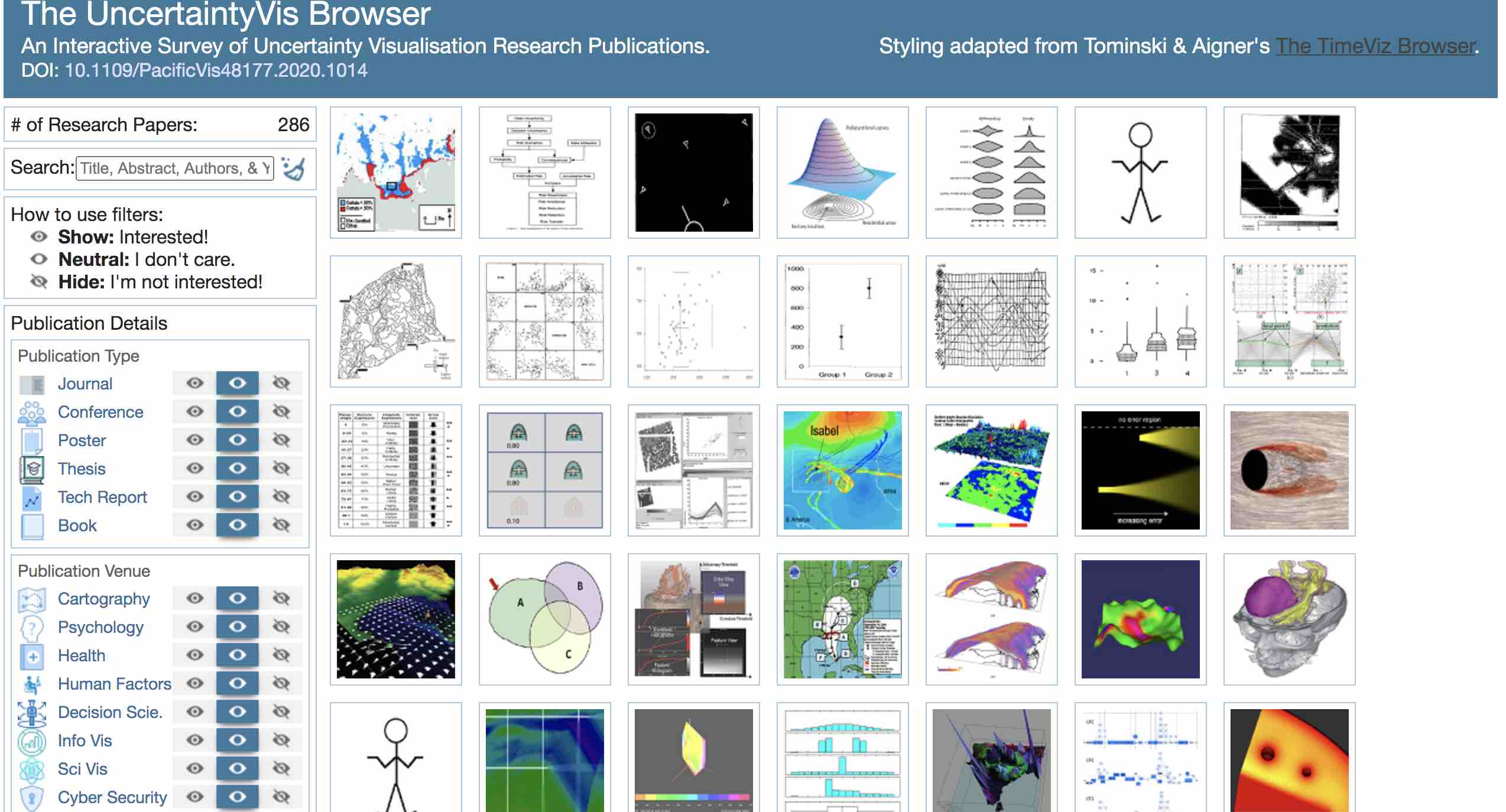}
        \end{minipage}%
        \begin{minipage}[t]{0.237\linewidth}
            \centering
            \includegraphics[width=1.7in]{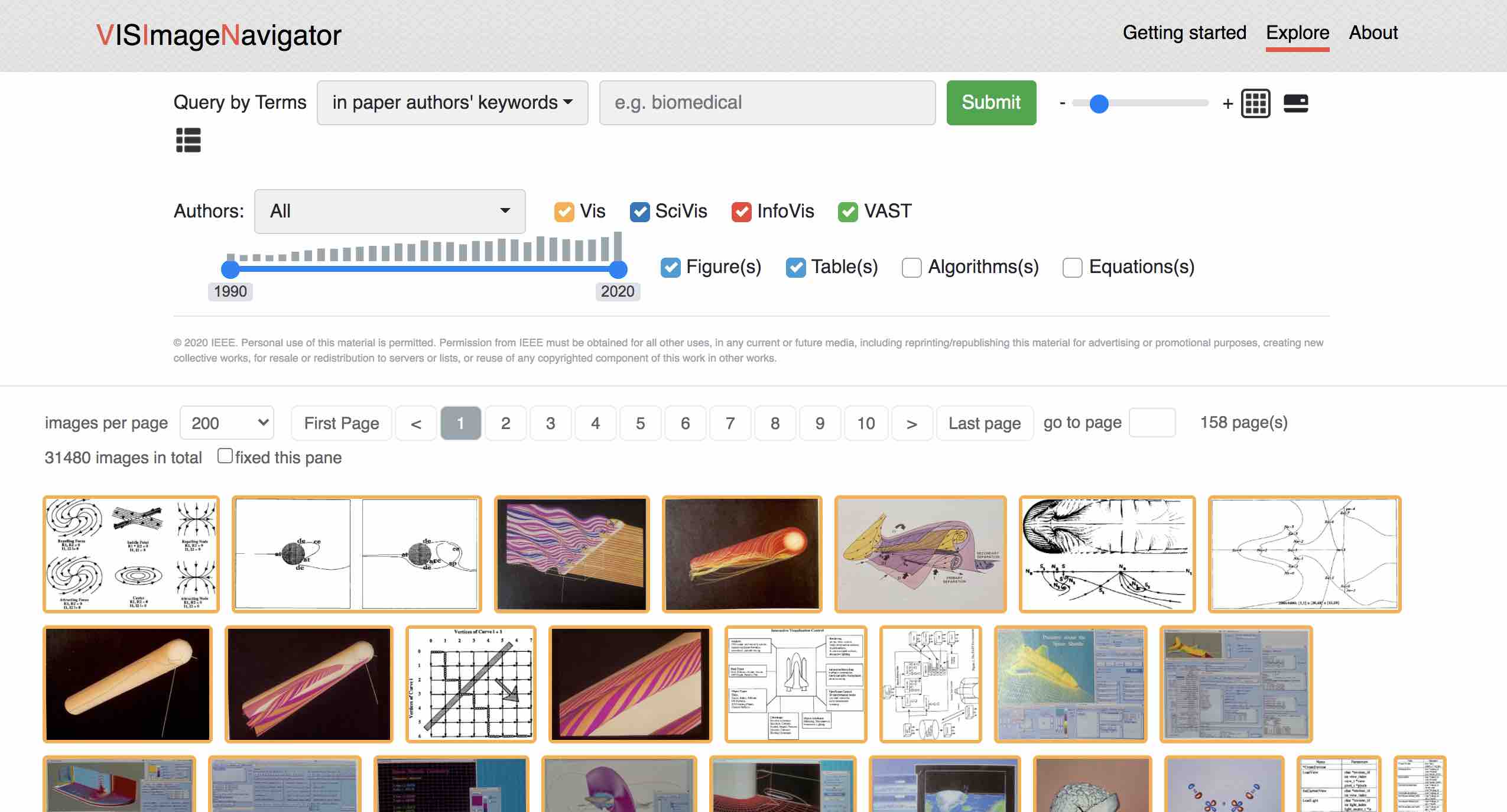}
        \end{minipage}%
    }%
    \centering
    \caption{Thumbnail images of Image Collection Browser web pages.  They are presented in the same chronological order as Table~\ref{tab:image}.}
    \label{fig:image}
\end{figure*}

A number of visualization survey papers have been published that also assemble an online collection of images related to the theme the survey covers. The list of web sites accompanying image collection browsers are summarized in Table~\ref{tab:image}.

\textbf{The Search Process:}
We came across these survey papers during a multi-year search process for visualization survey papers and state-of-the-art reports.
This search process was undertaken while we were writing the Survey of Surveys (SoS) in Information Visualization~\cite{McNabb2017c}.
During our search for survey papers in information visualization and visual analytics, which is described in great detail by McNabb and Laramee~\cite{McNabb2017c}, we made special note of all the surveys with accompanying image collection browsers.
After our initial findings based on publishing the SoS, we then performed two further searches for each research paper in this category: the recursive search and the forward looking search described previously.
\textbf{Inclusion Criteria:}
We include visualization survey papers and state-of-the-art reports that offer a collection of refereed visualization images that facilitates both search and comparison of special sub-topics in visualization and visual analytics.
Table~\ref{tab:image} and Figure~\ref{fig:image} provide an overview of image collection browsers.
One example of such a survey paper with an image collection browser is from Schulz \cite{Schulz2011} which features a collection of over 300 images that exemplify tree visualization.  Another excellent example of this is presented by  Aigner et al. \cite{Aigner2011a} \cite{Aigner2011} which hosts an online collection of over 100 images on the theme of time-oriented visualization. Kehrer and Hauser present an online collection of visualization images on the topic of multivariate  and multifaceted scientific data \cite{Kehrer2013a}. The collection features over 160 images related to this topic from refereed sources.

Kucher and Kerren collect a large, online collection of refereed data-visualization images \cite{Kucher2015}. Their collection holds over 470 images. Kerren et al. present an advanced image browser related to biological visualization techniques \cite{Kerren2017a}. This image browser has a special interactive feature that shows citations in the form of a graph with over 140 images. Kucher et al. present another survey paper on the state-of-the-art in sentiment visualization \cite{Kucher2018a}. This is also an advanced image browser that features a valuable collection of meta-data shown when clicking on an image. It holds over 160 peer-reviewed visualization images. Chatzimparmpas et al. collect 200 refereed images on the topic of building trustworthy machine learning methods using visualization \cite{Chatzimparmpas2020}. The image browser also offers a detailed set of meta-data for each image.

Dumas et al. developed an online collection of visualization images focussing on finance \cite{Dumas2014}. The image browser supports a number of filtering  options and stores over 85 peer-reviewed images.

Lu et al. collect an archive of online images related to the topic of predictive visual analytics \cite{Lu2017a}. A number of interactive filtering methods are available for over 70 images. Nobre et al. showcase both a collection of images and guidance on visual designs related to multivariate network visualization techniques\cite{Nobre2019b}. Jena et al. present an impressive visualization image browser with advanced filtering options for over 280 images\cite{Jena2020} all of which are dedicated to the topic of uncertainty visualization.

Chen et al.~\cite{chen2020vis30k} have the most complete collection
of the image data (figures, tables, equations, and algorithms) of
all three conference tracks of IEEE VIS in its 31 years of history. This dataset is also cross-linked to vispubdata~\cite{Isenberg2017b} so one can find images by keyword search. Their open-source models~\cite{ling2020deeppapercomposer} can make the subsequent data collection easier with minimum human intervention.

\paragraph*{\textbf{Survey Papers with an Accompanying SurVis Web Page}}\label{sec:Survis}

\begin{table*}[!tb]
    \caption{SurVis References: A summary of the SurVis web pages described in Section~\ref{sec:Survis}. These papers feature a related SurVis web page.See Figure~\ref{fig:Survis} for images of the SurVis collection web pages}
    \centering

    \resizebox{\textwidth}{!}{
        \begin{tabulary}{\textwidth}{|l|l|}
            \hline
            \textbf{Visualization Topic}  &  \textbf{SurVis URL}  \\ \hline
            Beck et al., \textbf{Dynamic Graph Visualization} \cite{Beck2014} & \url{http://dynamicgraphs.fbeck.com/}   \\ \hline
            Isaacs et al., \textbf{Performance Visualization} \cite{Isaacs2014} & \url{http://hdc.cs.arizona.edu/people/kisaacs/STAR/}   \\ \hline
            Vehlow et al., \textbf{Group Structures in Graphs}\cite{Vehlow2015} & \url{http://groups-in-graphs.corinna-vehlow.com/}   \\ \hline

            Nusrat and Kobourov, \textbf{ Cartograms} \cite{Nusrat2016a} & \url{http://cartogram.cs.arizona.edu/survis-cartogram/}   \\ \hline
            Liu et al., \textbf{Visualizing High-Dimensional Data: Advances in the Past Decade} \cite{Liu2017} & \url{http://www.sci.utah.edu/~shusenl/highDimSurvey/website/}   \\ \hline
            Federico et al.,\textbf{PaperViz.org} \cite{Federico2016c} & \url{http://ieg.ifs.tuwien.ac.at/~federico/LiPatVis/}   \\ \hline
            Beck and Weiskopf, \textbf{Sparklines Literature} \cite{Beck2017a} & \url{http://sparklines-literature.fbeck.com/}   \\ \hline
            Windhager et al.,\textbf{ Collectionvis.org} \cite{Windhager2019a} & \url{http://collectionvis.org/}   \\ \hline
        \end{tabulary}
    }
    \label{tab:SurVis}
\end{table*}

\begin{figure*}[!tb]
    \centering
    \subfigure{
        \begin{minipage}[t]{0.237\linewidth}
            \centering
            \includegraphics[width=2in]{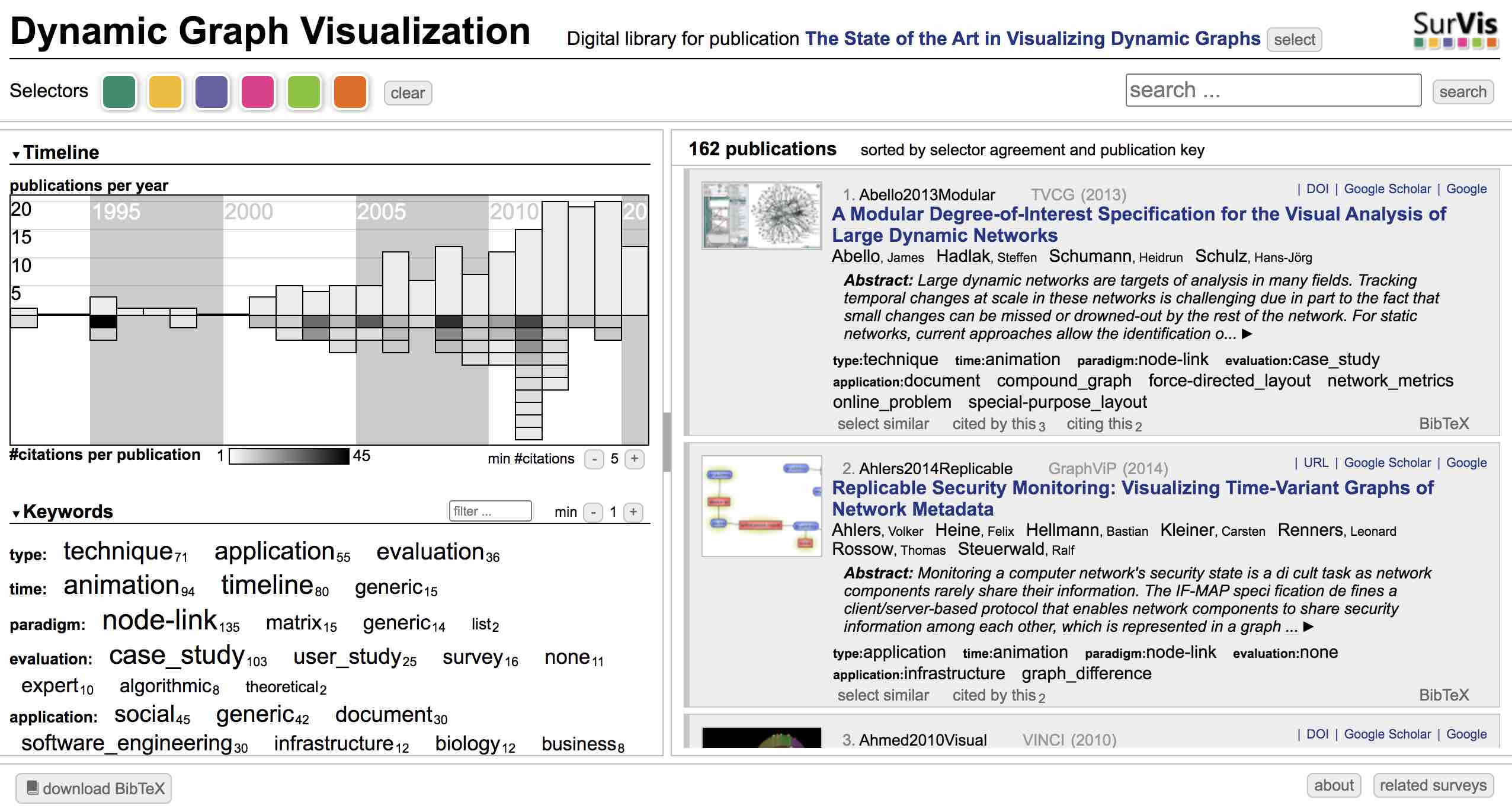}
        \end{minipage}%
        \begin{minipage}[t]{0.237\linewidth}
            \centering
            \includegraphics[width=2in]{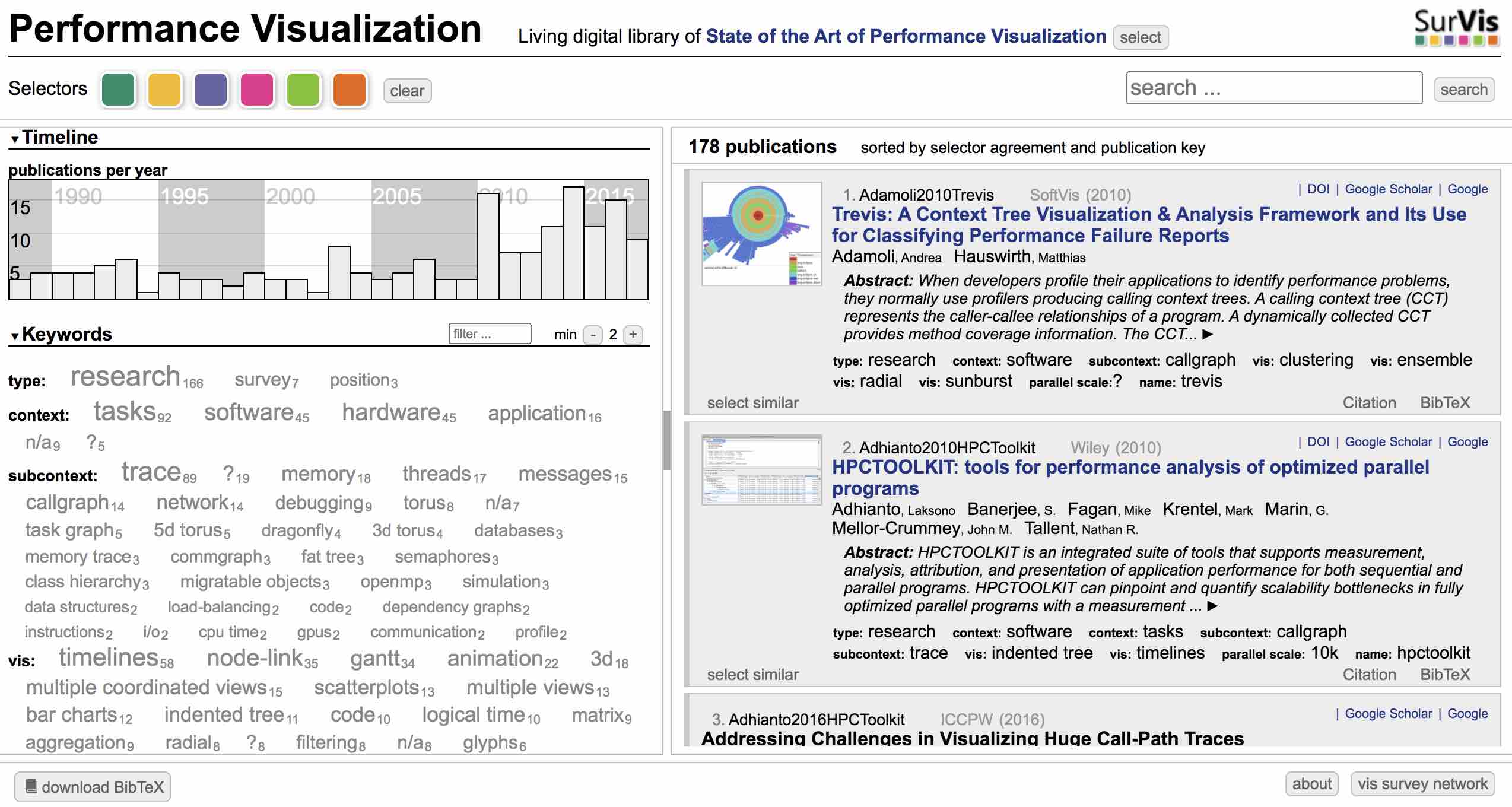}
        \end{minipage}%
        \begin{minipage}[t]{0.237\linewidth}
            \centering
            \includegraphics[width=2in]{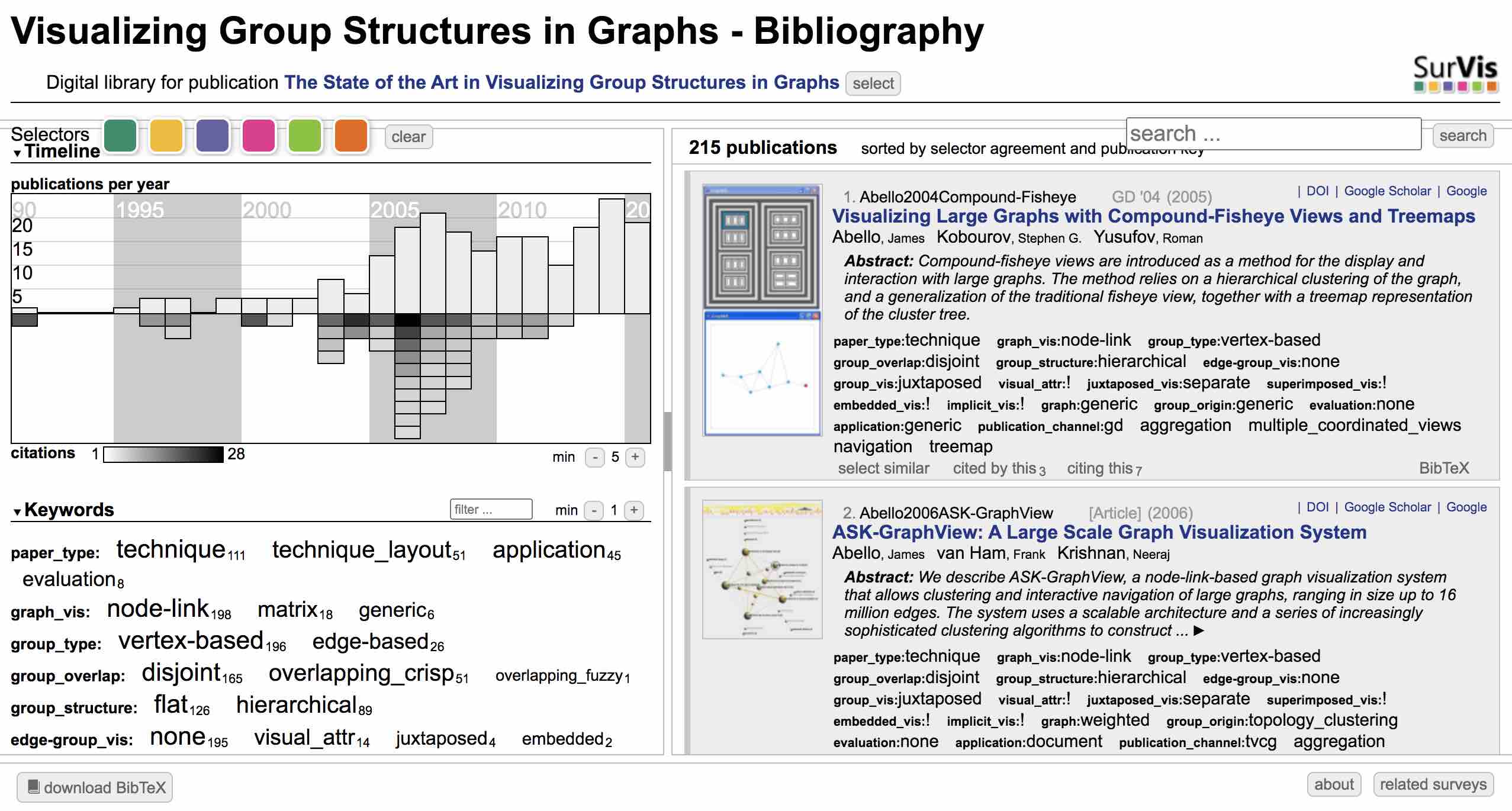}
        \end{minipage}%
        \begin{minipage}[t]{0.237\linewidth}
            \centering
            \includegraphics[width=2in]{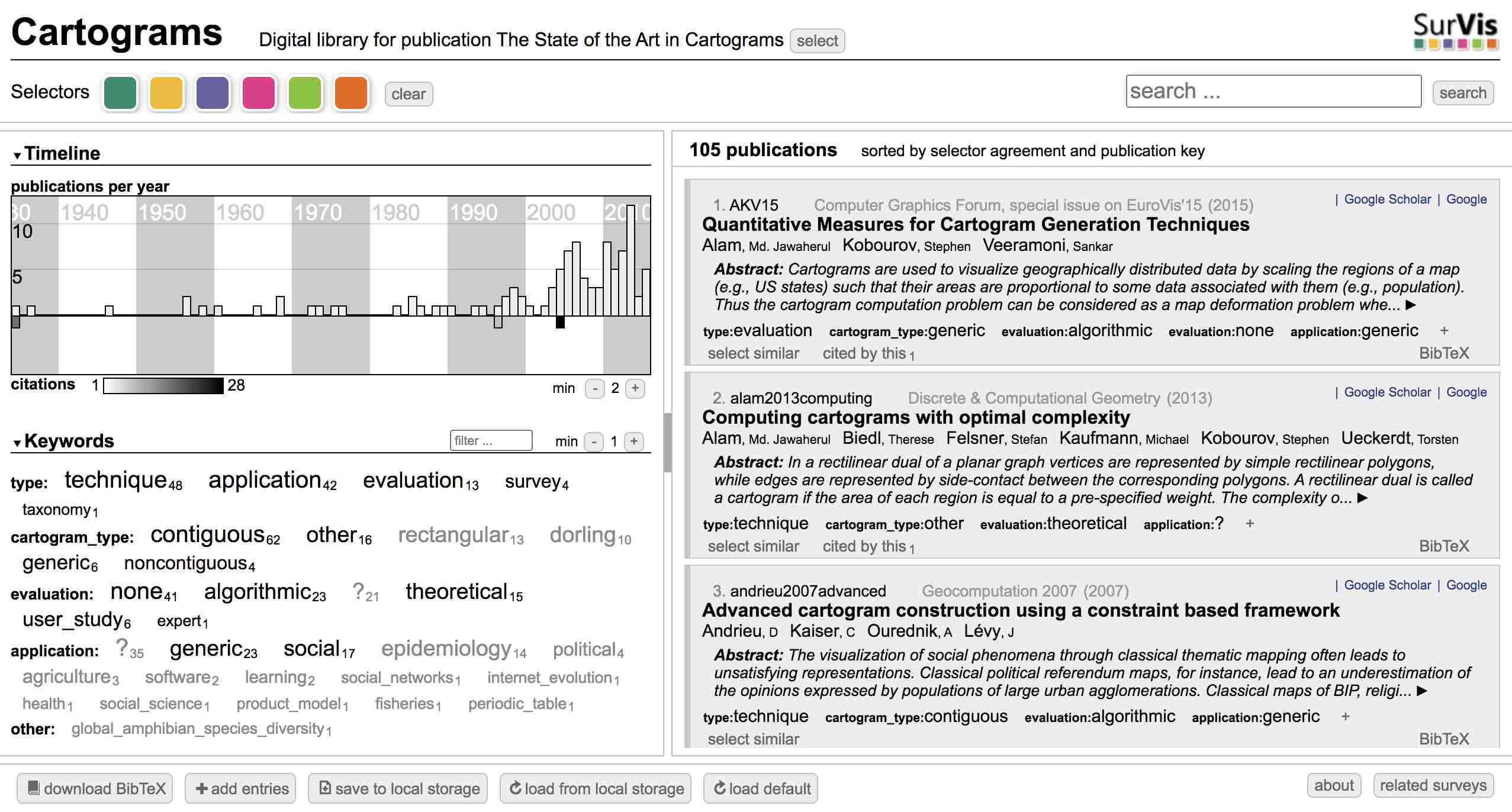}
        \end{minipage}%
    }%

    \subfigure{
        \begin{minipage}[t]{0.237\linewidth}
            \centering
            \includegraphics[width=2in]{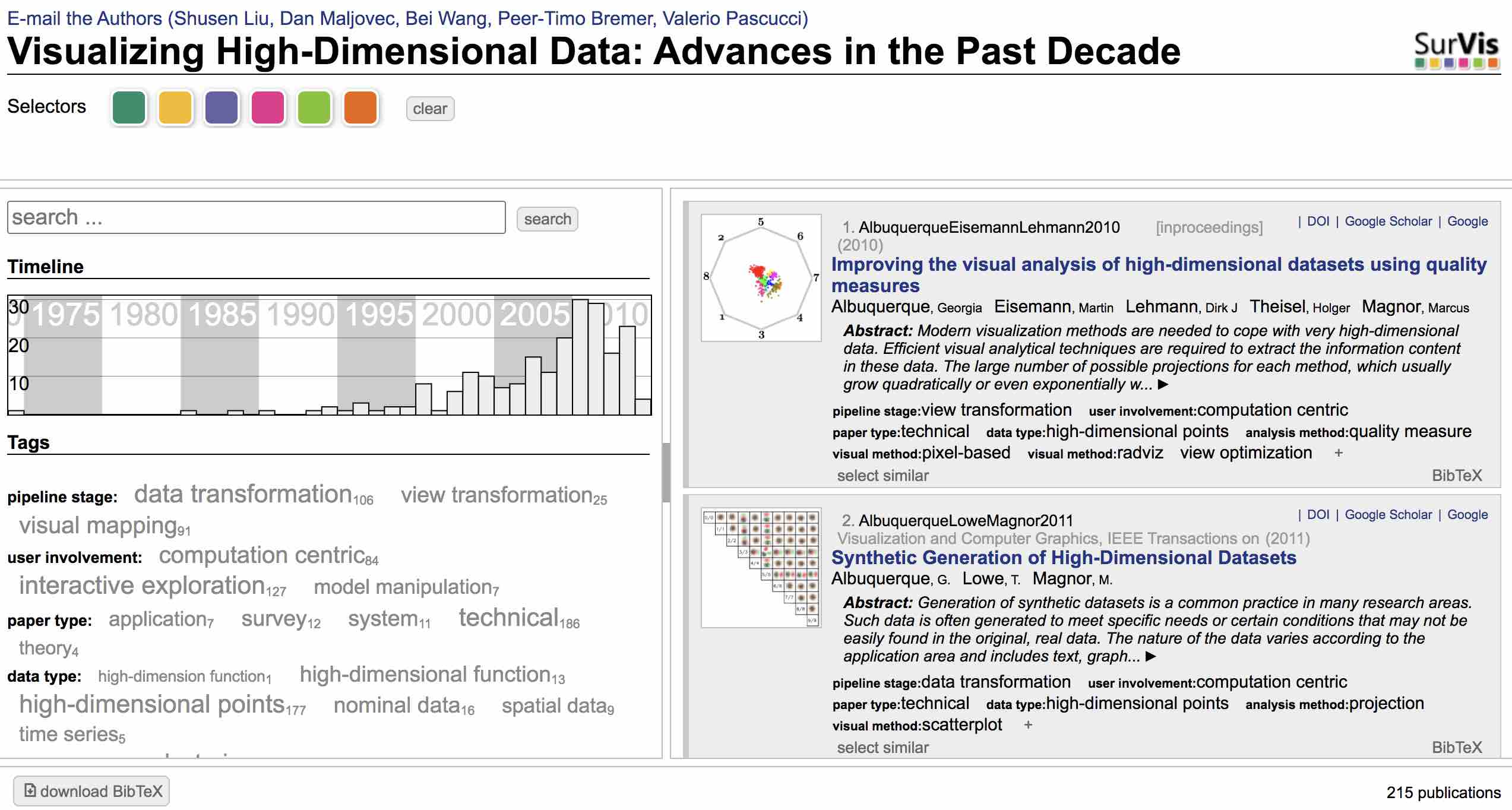}
        \end{minipage}%
        \begin{minipage}[t]{0.237\linewidth}
            \centering
            \includegraphics[width=2in]{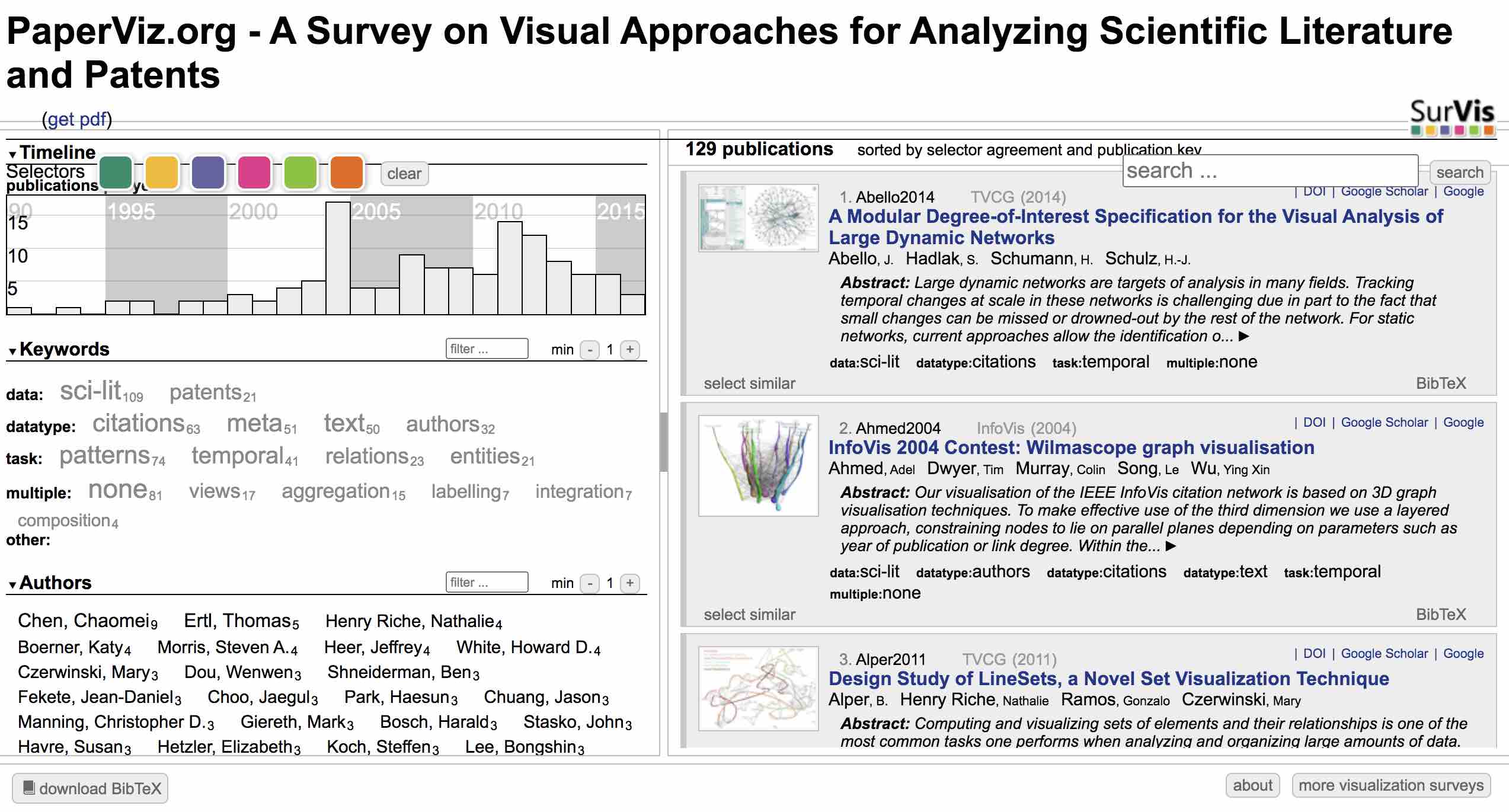}
        \end{minipage}%
        \begin{minipage}[t]{0.237\linewidth}
            \centering
            \includegraphics[width=2in]{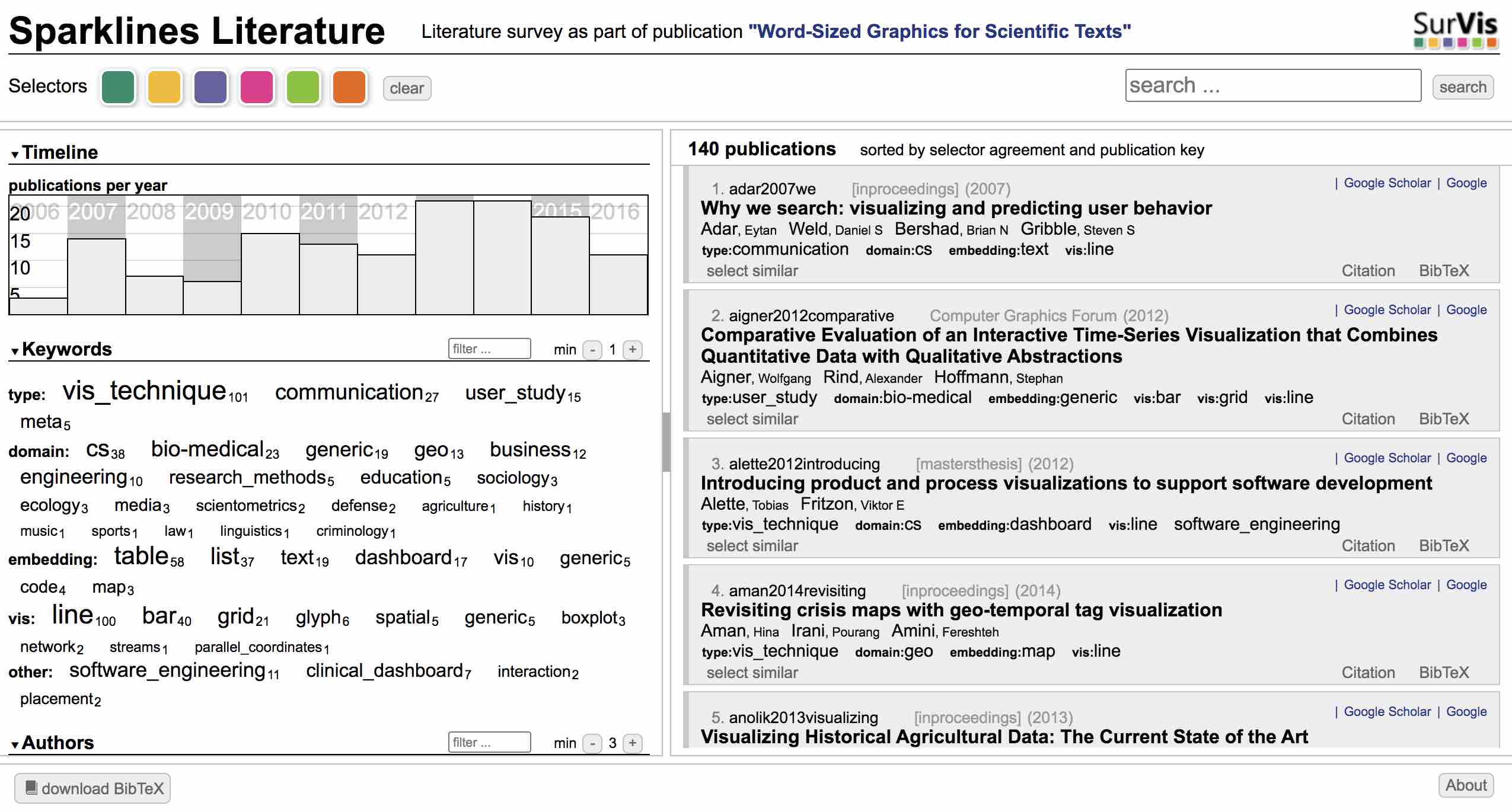}
        \end{minipage}%
        \begin{minipage}[t]{0.237\linewidth}
            \centering
            \includegraphics[width=2in]{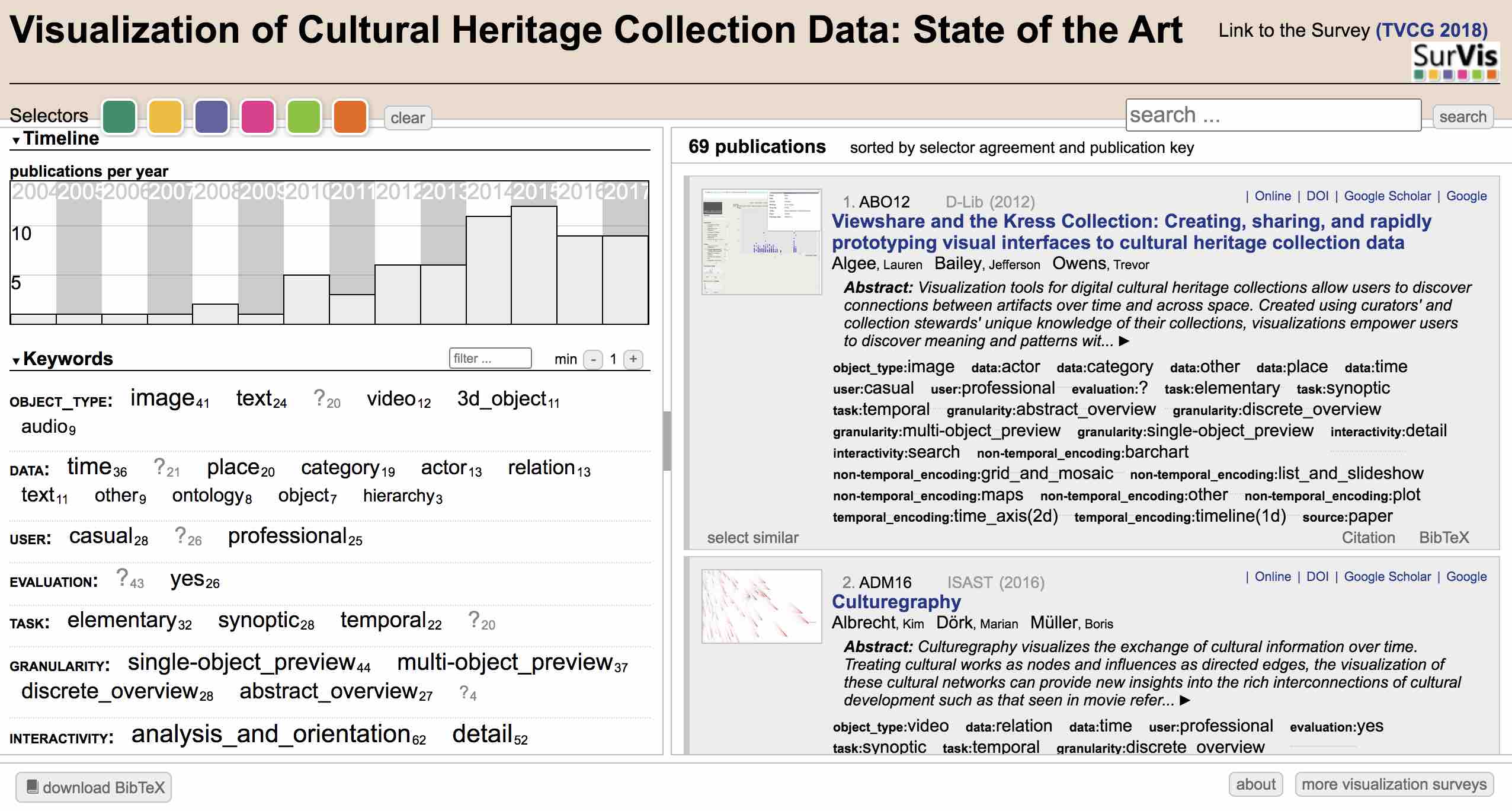}
        \end{minipage}%
    }%

    \centering
    \caption{Thumbnail images of SurVis web pages.  They are presented in the same order as Table~\ref{tab:SurVis}}
    \label{fig:Survis}
\end{figure*}

SurVis \cite{Beck2016b} is a flexible visual analytics tool used to structure and present the list of references in surveys. The list of web sites with SurVis Web Page are summarized in Table~\ref{tab:SurVis} and Figure~\ref{fig:Survis}.

The SurVis interface is divided into two main views. An overview of the literature collection and a detailed list of references accompanied by an image. In the overview, the collection is presented using a timeline that provides a yearly chronology of the publication collection. The barchart is integrated into the timeline to summarize the number of publications in each year. The rest of the overview leverages selectable word-sized sparkline visualizations embedded in word clouds to visualize the publications metadata and clusters. The selector mechanism enables linkage between the two views.

\textbf{The Search Process:}
We discovered several survey papers with accompanying SurVis web pages during a multi-year search process for visualization survey papers and state-of-the-art reports.
The initial search process was carried out while we were writing the Survey of Surveys (SoS) in Information Visualization and Visual Analytics~\cite{McNabb2017c}.
During our search for survey papers in information visualization and visual analytics we made special note of all the surveys with accompanying SurVis web pages.
After our initial findings based on publishing the SoS, we then performed a
forward looking search based on the original SurVis publication~\cite{Beck2016b}.
We used Google's ``cited by" feature to find all research papers citing the
SurVis paper.
\textbf{Inclusion Criteria:}
We include visualization and visual analytics survey papers and state-of-the-art reports that offer an accompanying SurVis web page.

Several surveys incorporate SurVis to structure and visually analyze a literature collection. Beck et al.\cite{Beck2014} use SurVis to systematically derive a hierarchical taxonomy of dynamic visualizations. Isaacs et al. \cite{Isaacs2014} and Vehlow et al.\cite{Vehlow2015} use SurVis to exhibit their literature collection. The former presents visualization approaches that inform users on optimizing software performance. The latter presents approaches that explicitly depict group structures in graphs. Nusrat and Kobourov\cite{Nusrat2016a} use SurVis in their bibliographic analysis. Nusrat and Kobourov applied some modifications to SurVis to incorporate different cartogram types and applications. Liu et al.\cite{Liu2017} survey the approaches that visualize high-dimensional data. Federico et al. \cite{Federico2016c} review interactive analysis and visualization approaches of patents and scientific literature analysis. Beck and Weiskopf \cite{Beck2017a} present the state-of-the-art of embedding word-sized graphics within scientific texts. Windhager et al.\cite{Windhager2019a} also use SurVis to present their literature collection which focuses on visual interface design for cultural heritage collection data.

\paragraph*{\textbf{Surveys of Books}}
Books are traditionally overlooked when compiling state-of-the-art reports, however, they can contain a vast trove of information. A survey by Rees and Laramee addresses this by reporting on information visualization books \cite{Rees2019a}. In total, 41 books are reviewed totalling over 23,000 pages with a combined value of approximately \$3,600 USD. Books are classified according to the audience with recommendations provided for readers, along with an indication of how many pages are dedicated to each topic.

A number of web resources also list information visualization books.
\textbf{The Search Process:}
In order to find collections of information visualization books, we used search terms such as ``List of Information Visualization Books", ``Collections of Information Visualization Books", and ``Information Visualization Book Recommendations".  We also browsed the collection of web sites described in
Section~\ref{sec:websites}.
\textbf{Inclusion Criteria:}
We include the lists we found that were recent, less than two years old, and that listed at least 10 relevant books.

The team at Information is Beautiful has compiled a list of 73 books on information visualization and infographics with a very short description\cite{DataVisualizationSociety}. An extended list of 155 books also includes books covering programming tools\cite{Society}. Another list of 18 books has been compiled by Durcevic on the datapine blog along with a brief description of each book\cite{Sandra2019}. As part of their data visualization field guide, Tableau has compiled a list of 12 books including a few paragraphs of description \cite{Tableau}. Yet another list has been compiled by King from Solutions Review which includes 30 books and a single paragraph description of each \cite{TimothyKing}. The book recommendation website Goodreads lists books in the information visualization genre with a total of 54 books \cite{GoodReads}.

\section{Visualization-Focused Websites}
\label{sec:websites}

\begin{figure}[!tb]
    \centering
    \subfigure{
        \begin{minipage}[t]{0.23\linewidth}
            \centering
            \includegraphics[width=0.9in]{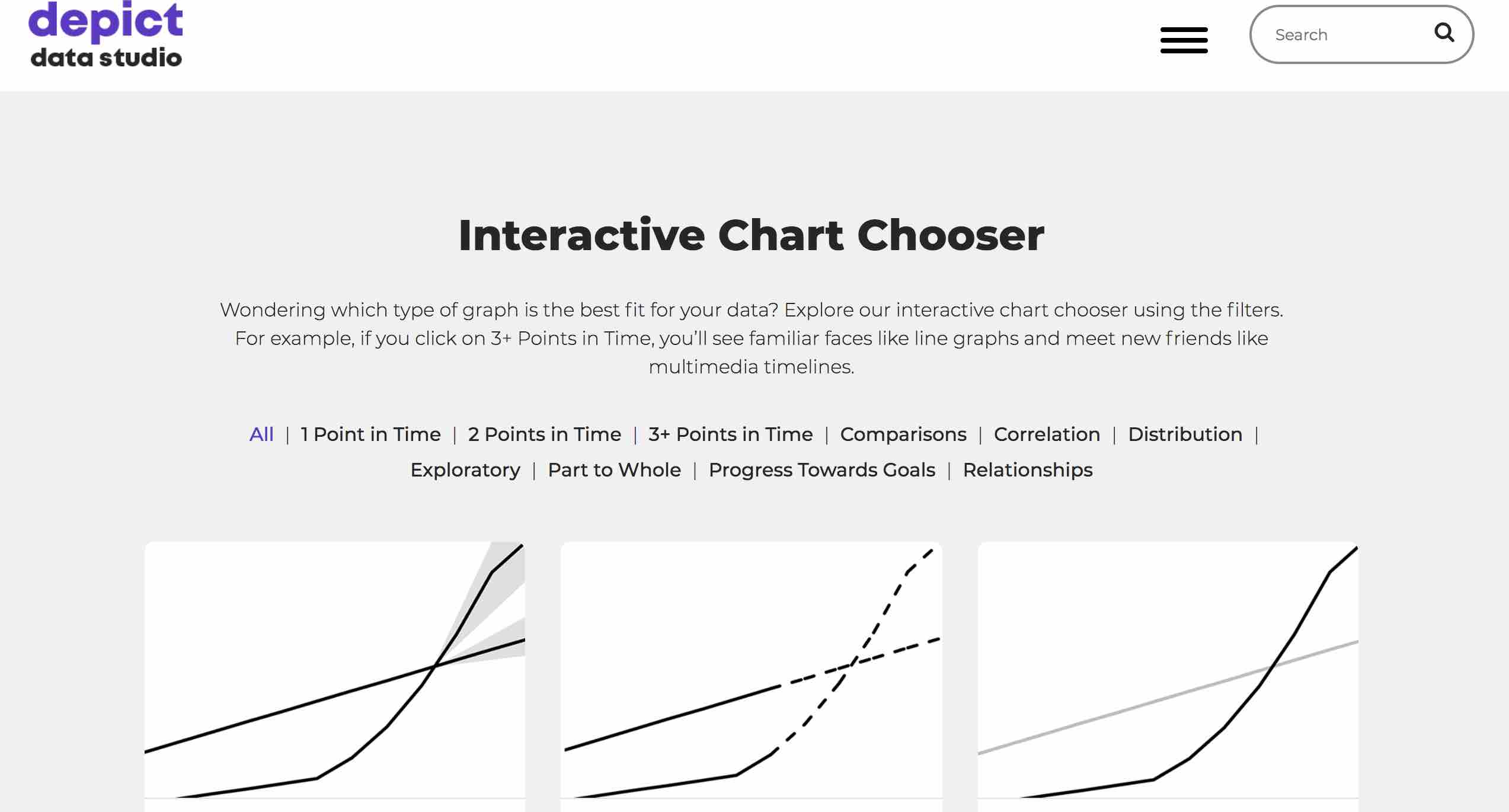}
        \end{minipage}%
        \begin{minipage}[t]{0.23\linewidth}
            \centering
            \includegraphics[width=0.9in]{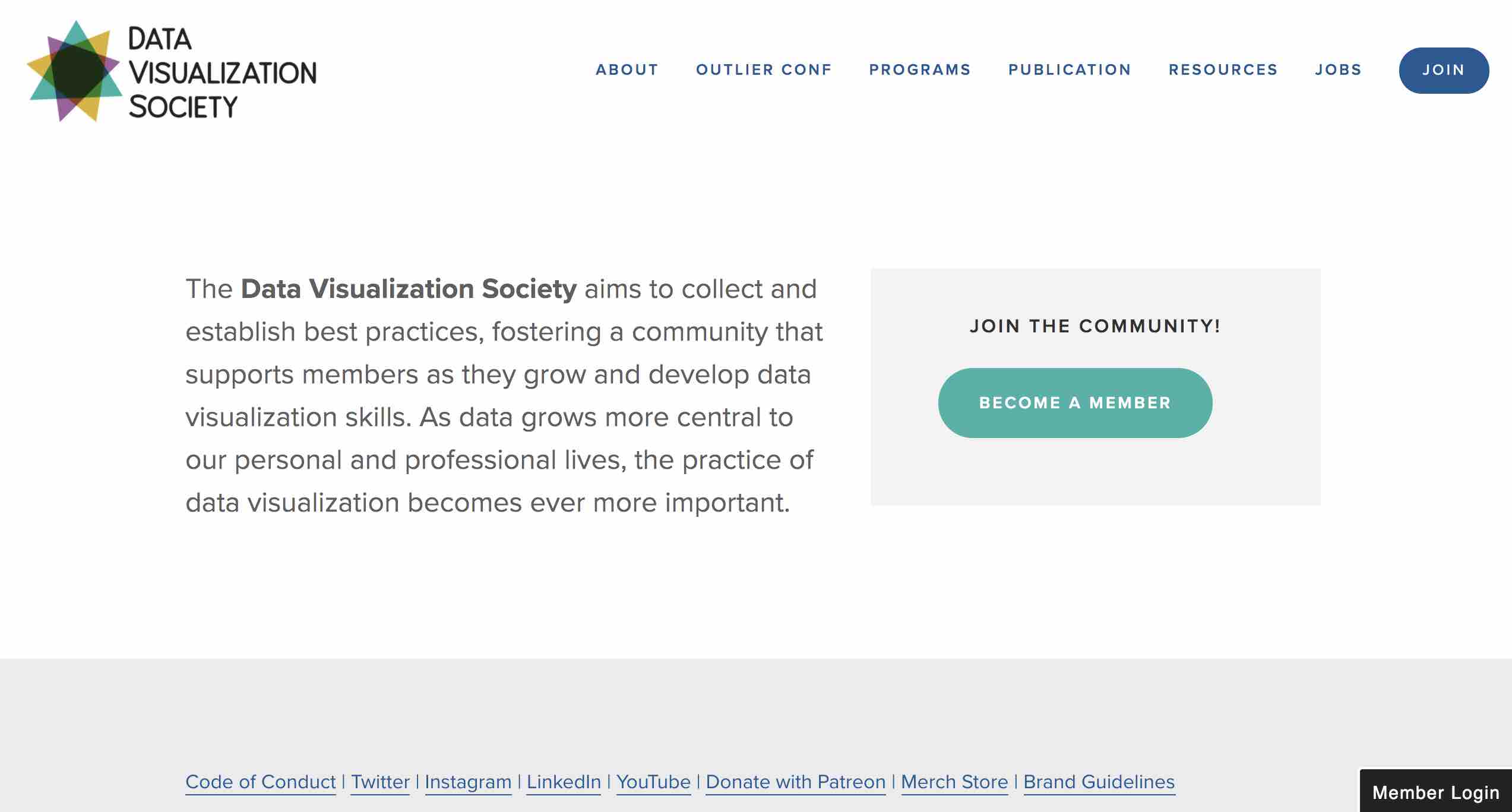}
        \end{minipage}%
        \begin{minipage}[t]{0.23\linewidth}
            \centering
            \includegraphics[width=0.9in]{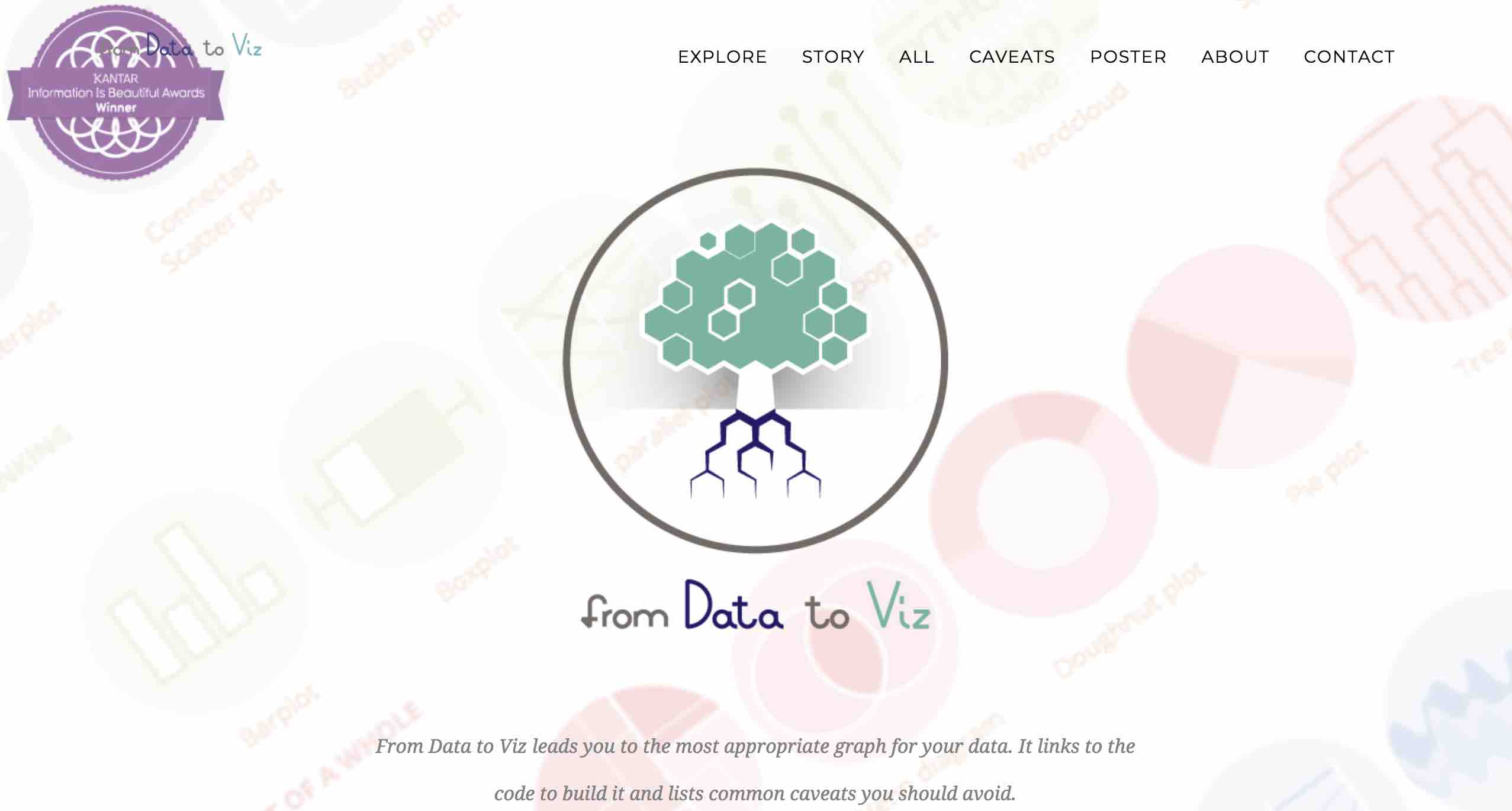}
        \end{minipage}%
        \begin{minipage}[t]{0.23\linewidth}
            \centering
            \includegraphics[width=0.9in]{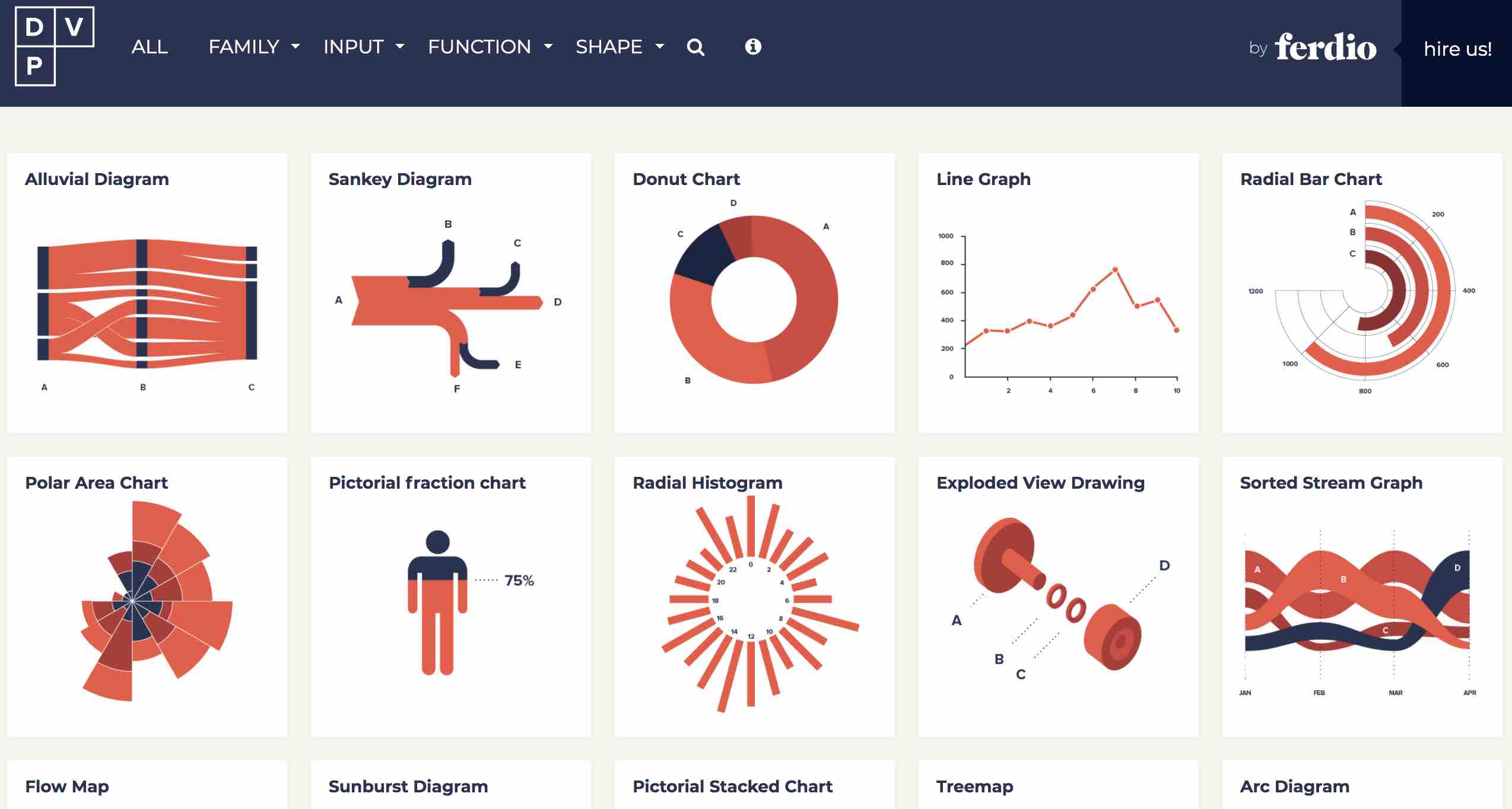}
        \end{minipage}%
    }%

    \subfigure{
        \begin{minipage}[t]{0.23\linewidth}
            \centering
            \includegraphics[width=0.9in]{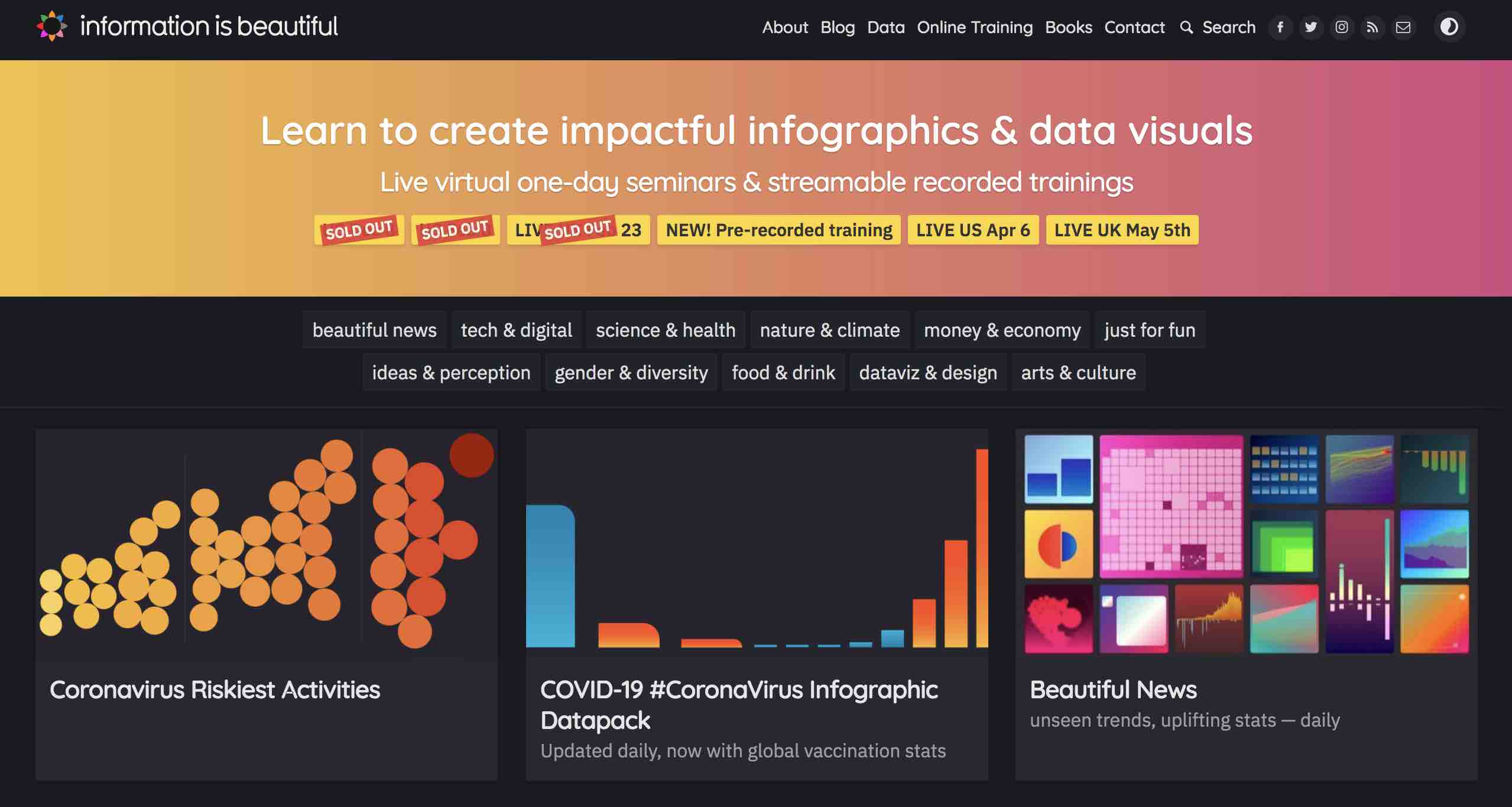}
        \end{minipage}%
        \begin{minipage}[t]{0.23\linewidth}
            \centering
            \includegraphics[width=0.9in]{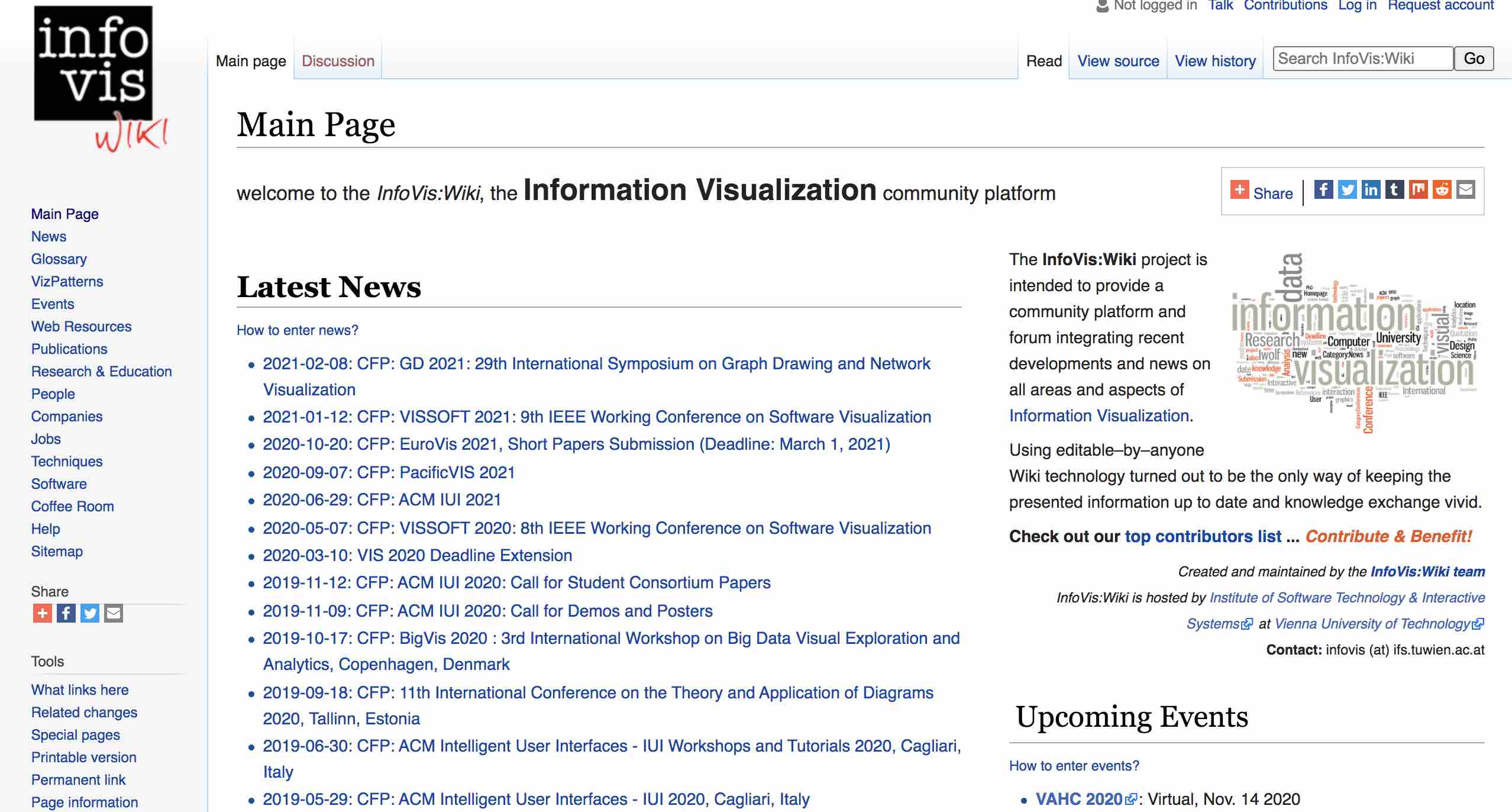}
        \end{minipage}%
        \begin{minipage}[t]{0.23\linewidth}
            \centering
            \includegraphics[width=0.9in]{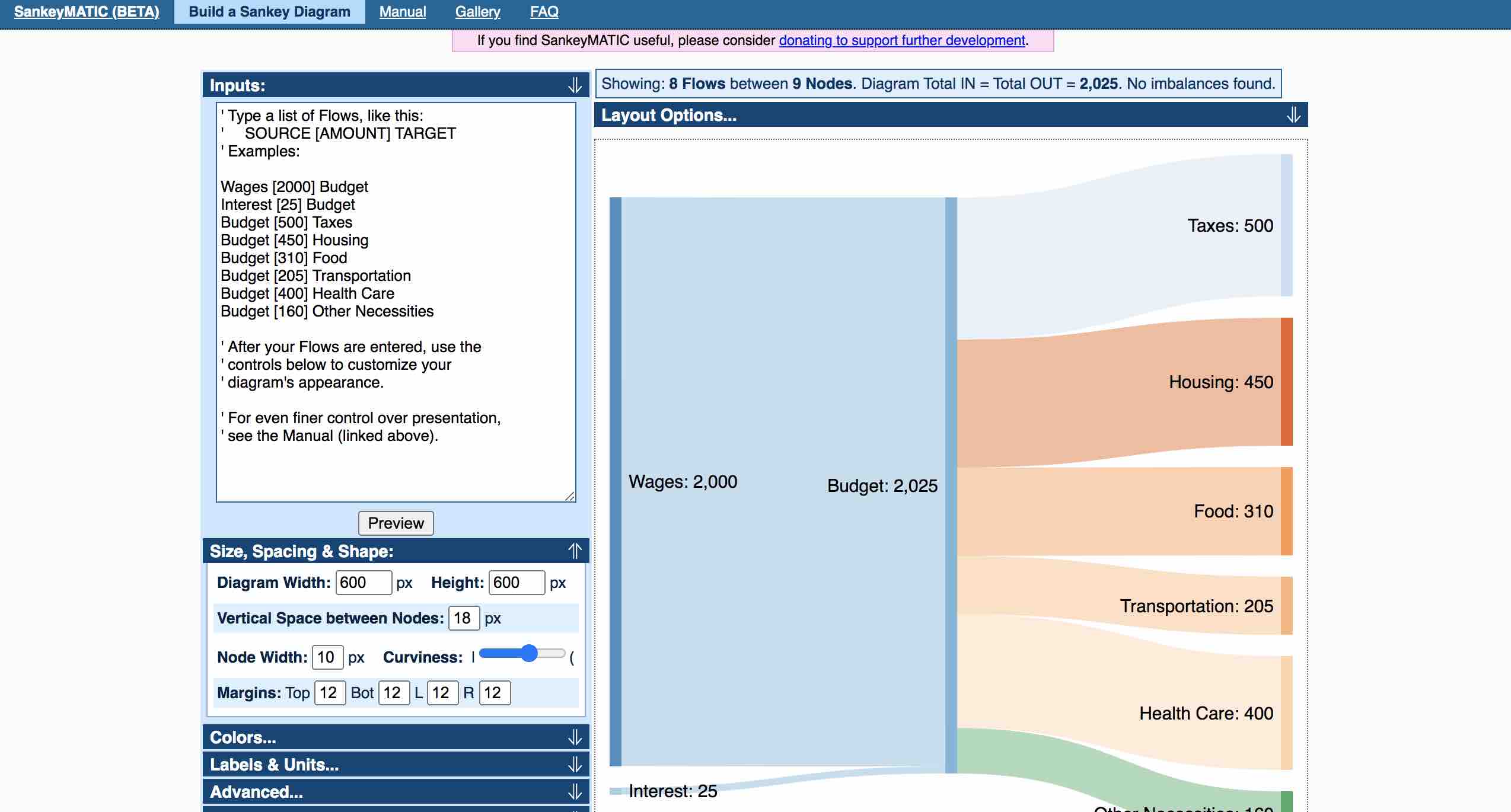}
        \end{minipage}%
        \begin{minipage}[t]{0.23\linewidth}
            \centering
            \includegraphics[width=0.9in]{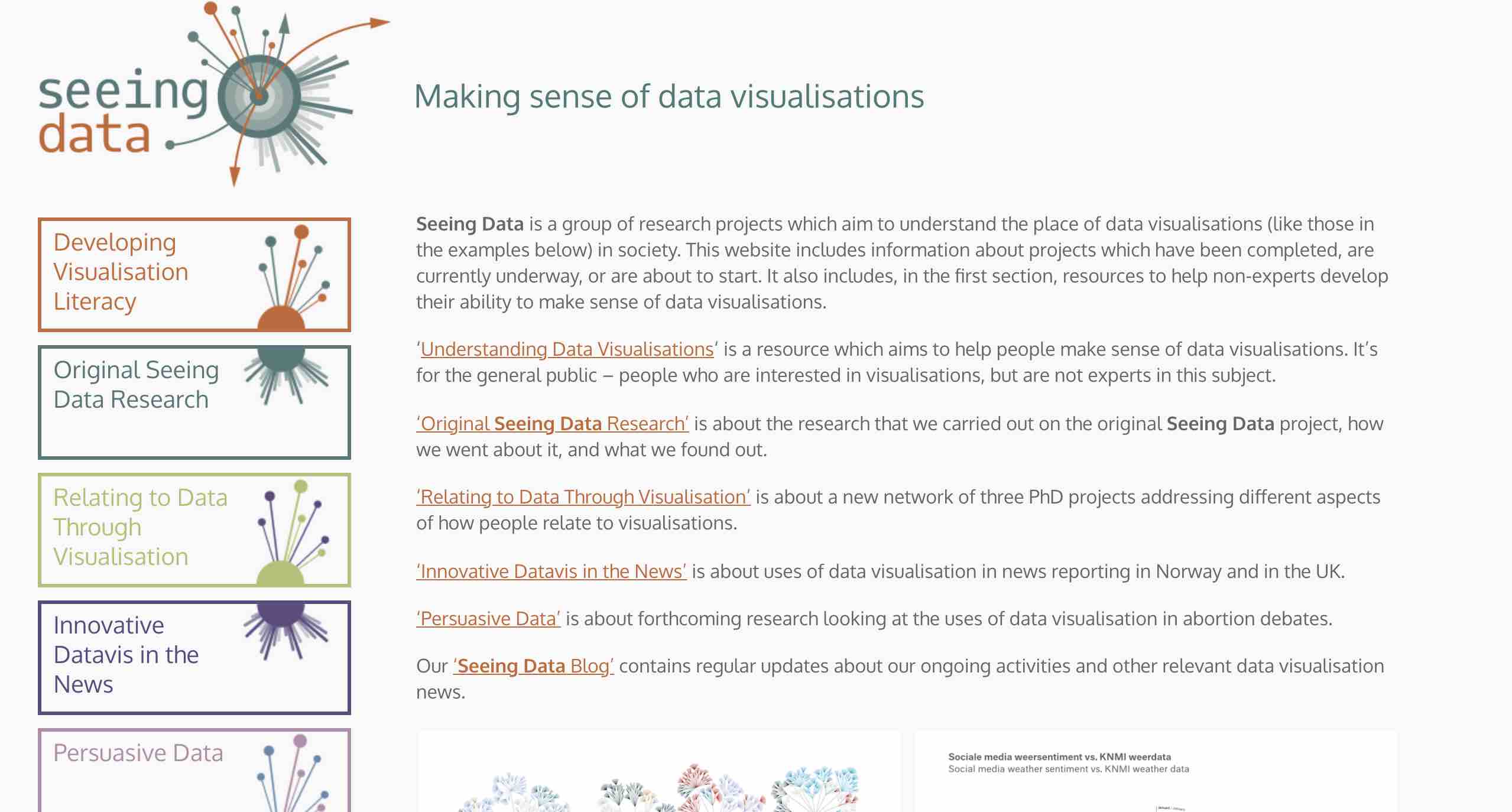}
        \end{minipage}%
    }%

    \subfigure{
        \begin{minipage}[t]{0.23\linewidth}
            \centering
            \includegraphics[width=0.9in]{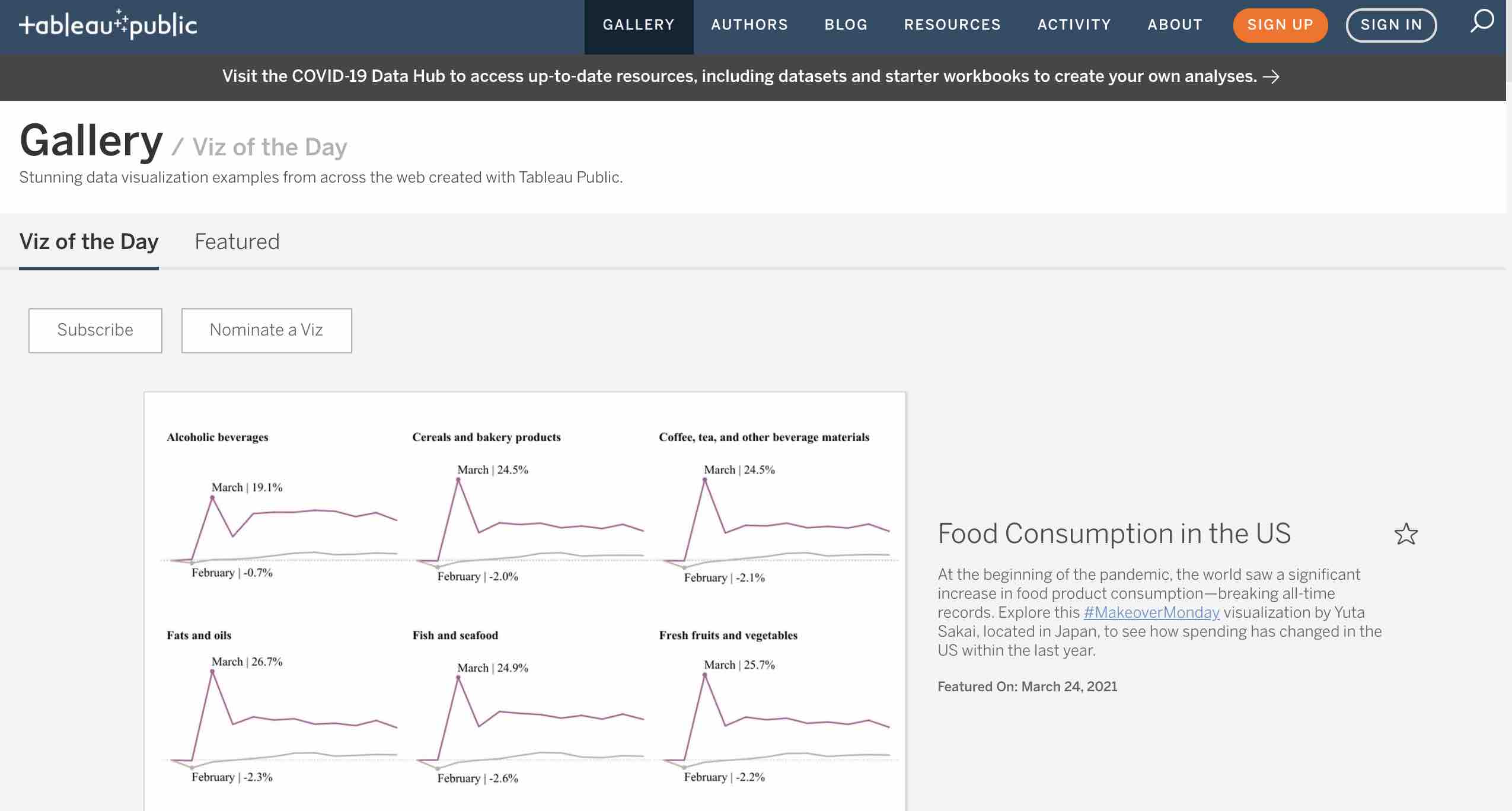}
        \end{minipage}%
        \begin{minipage}[t]{0.23\linewidth}
            \centering
            \includegraphics[width=0.9in]{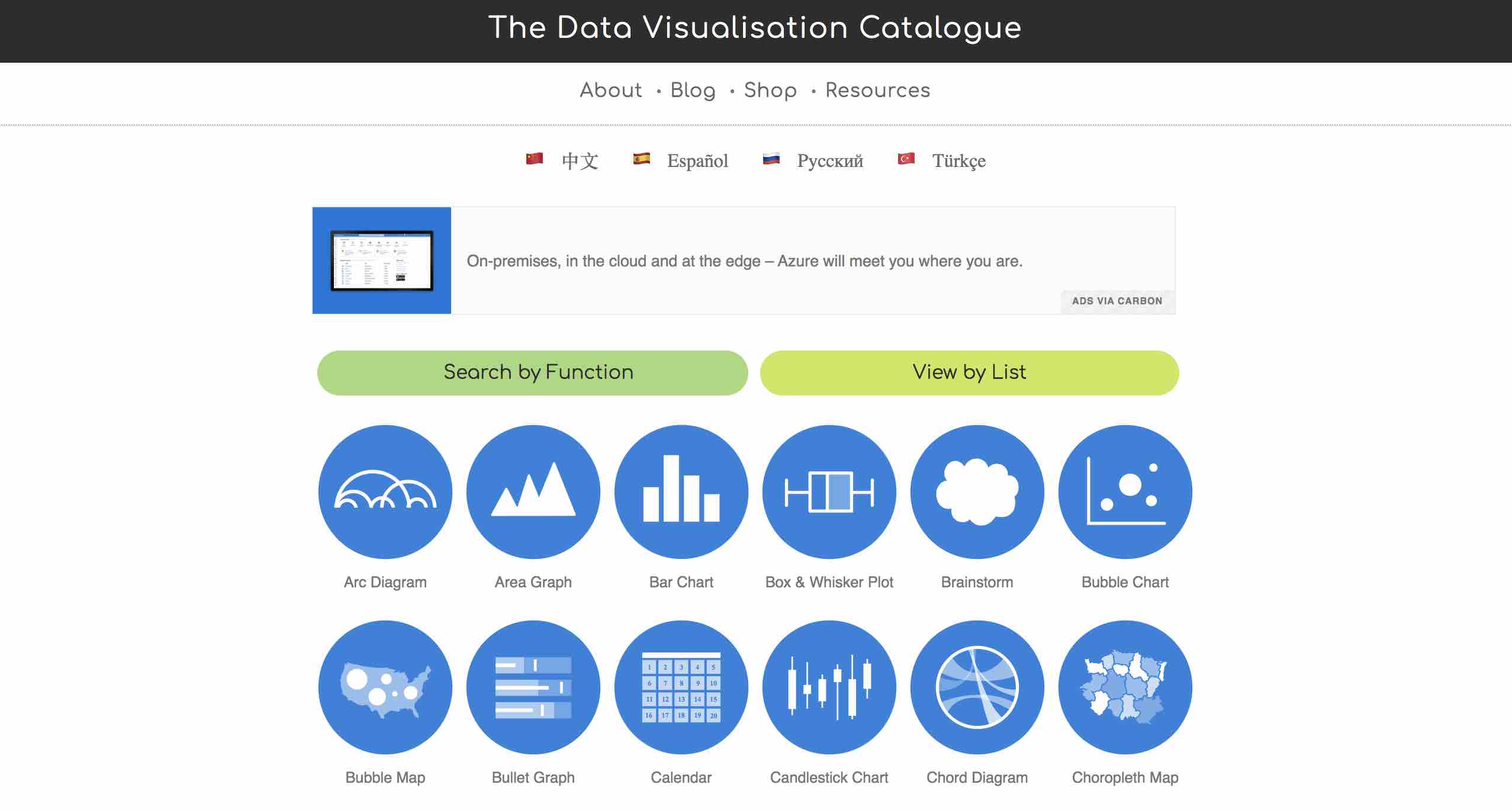}
        \end{minipage}%
        \begin{minipage}[t]{0.23\linewidth}
            \centering
            \includegraphics[width=0.9in]{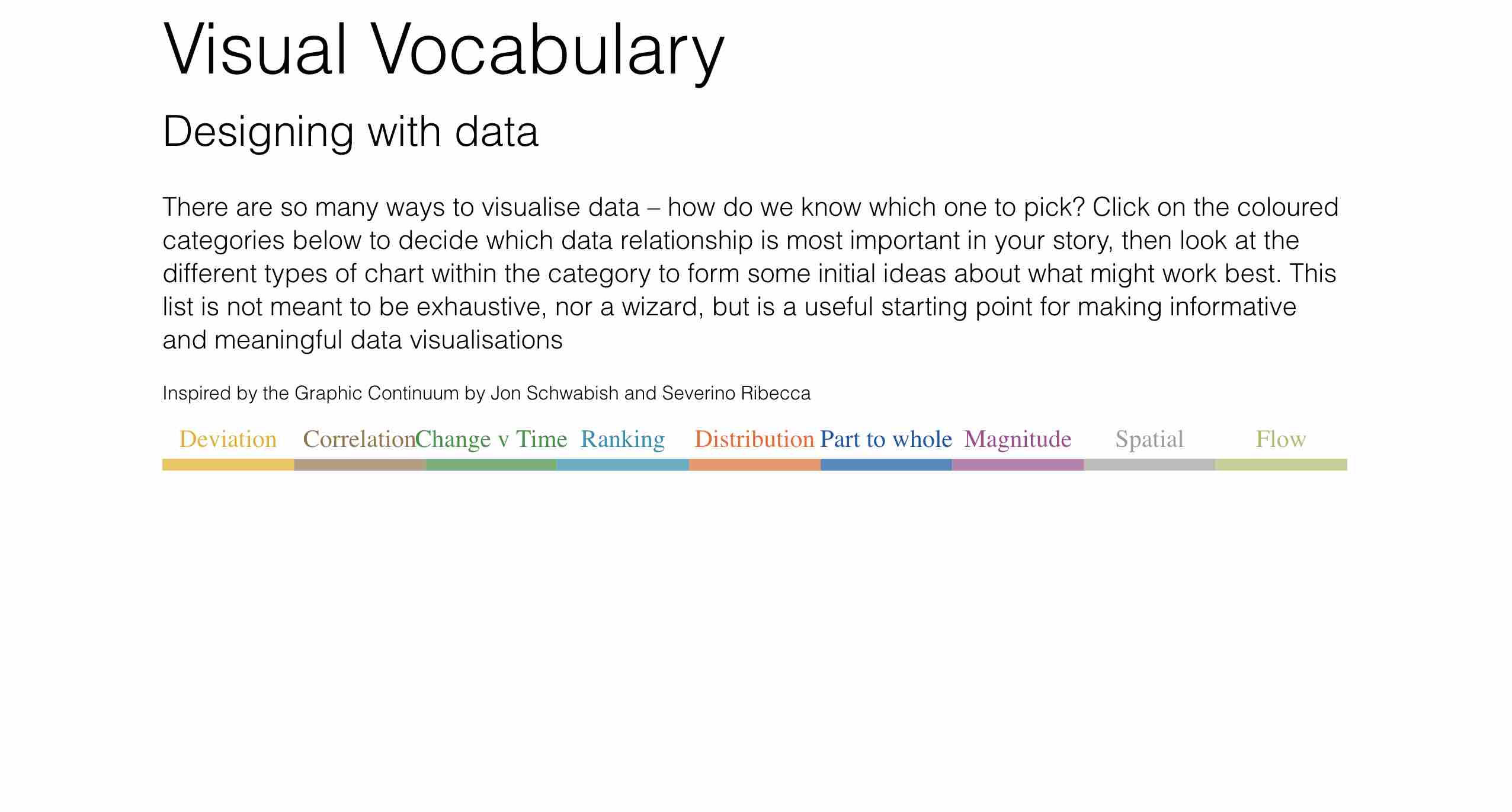}
        \end{minipage}%
        \begin{minipage}[t]{0.23\linewidth}
            \centering
            \includegraphics[width=0.9in]{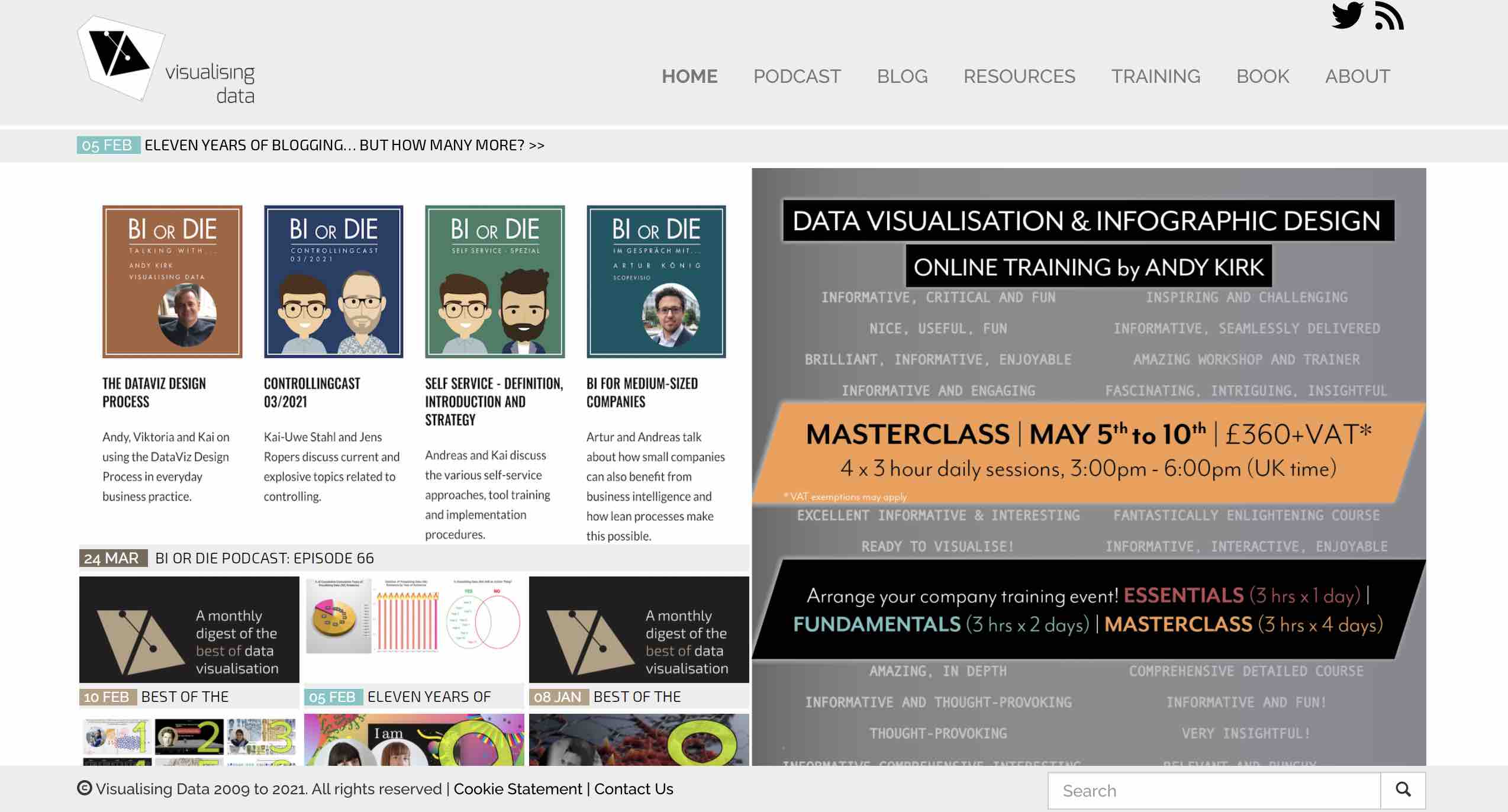}
        \end{minipage}%
    }%

    \subfigure{
        \begin{minipage}[t]{0.23\linewidth}
            \centering
            \includegraphics[width=0.9in]{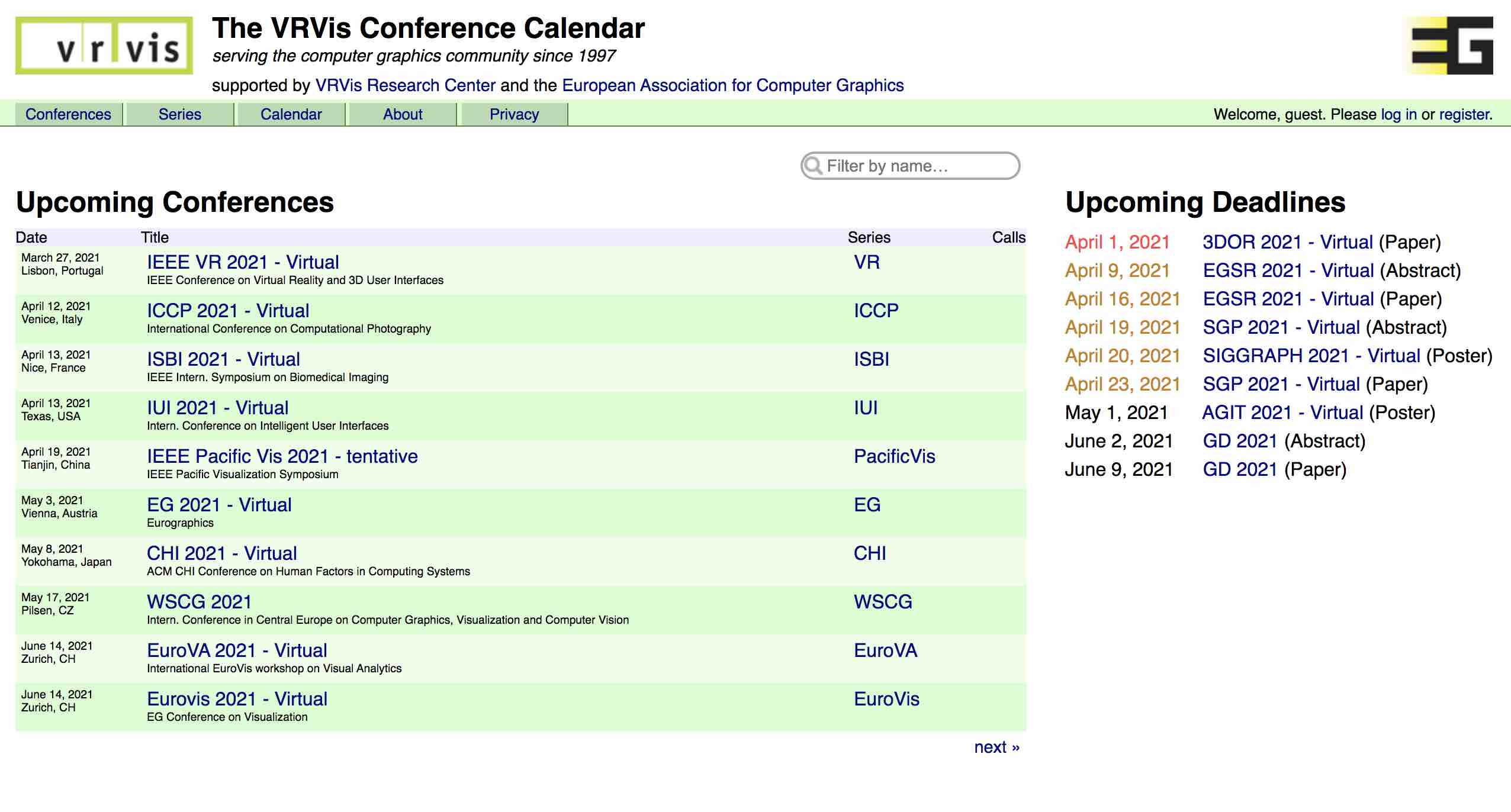}
        \end{minipage}%
        \begin{minipage}[t]{0.23\linewidth}
            \centering
            \includegraphics[width=0.9in]{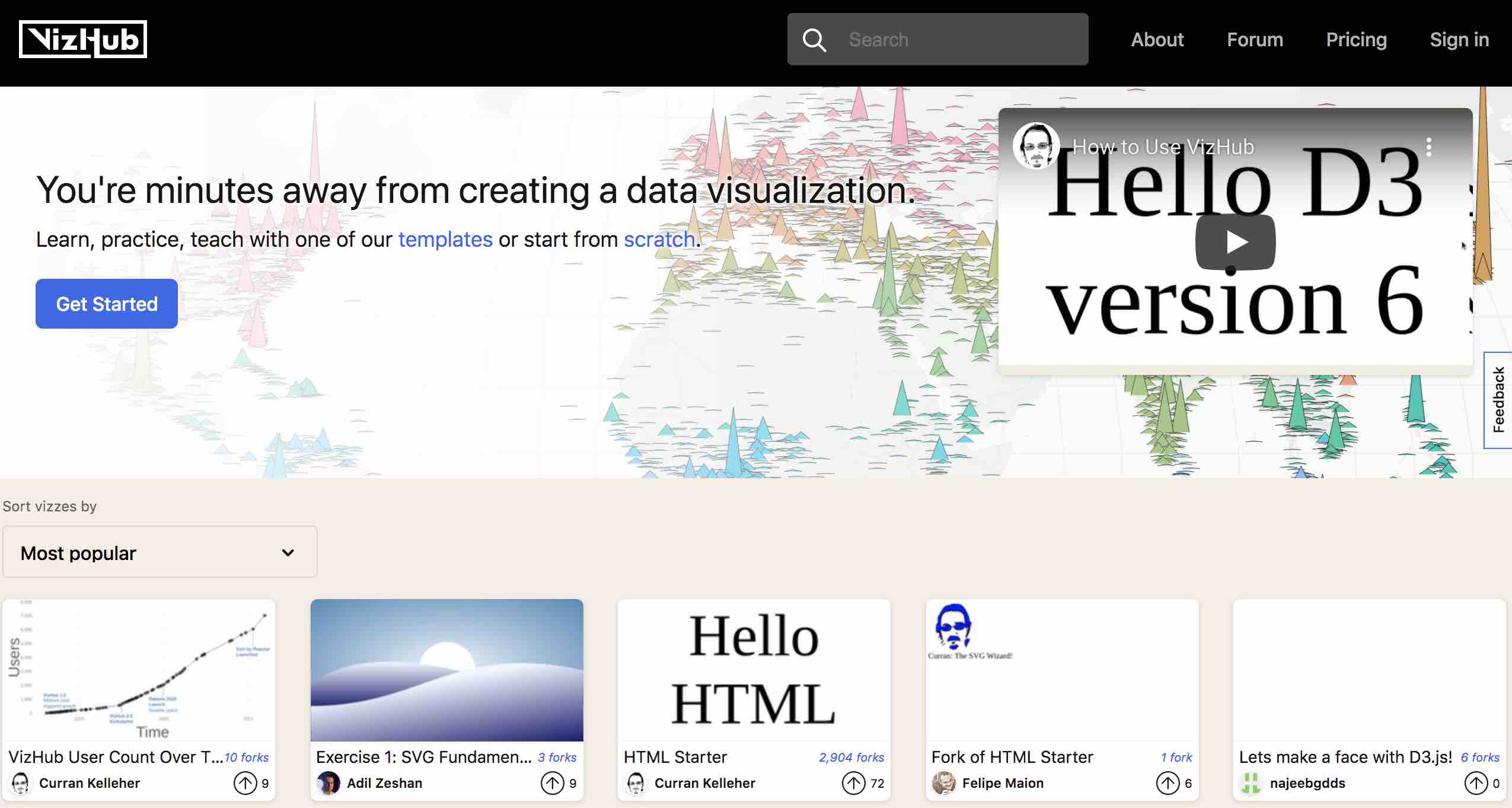}
        \end{minipage}%
        \begin{minipage}[t]{0.23\linewidth}
            \centering
            \includegraphics[width=0.9in]{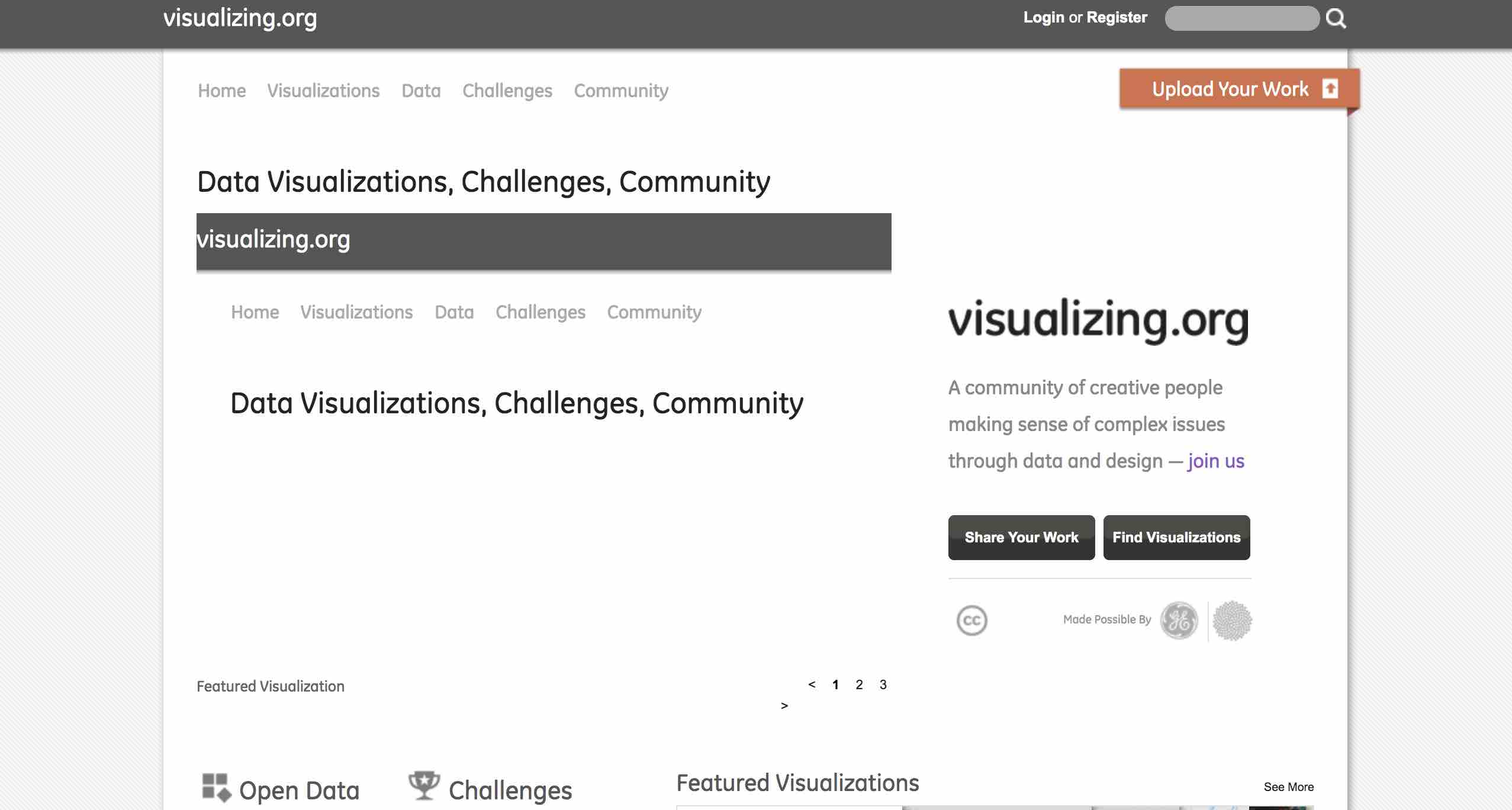}
        \end{minipage}%
        \begin{minipage}[t]{0.23\linewidth}
            \centering
            \includegraphics[width=0.9in]{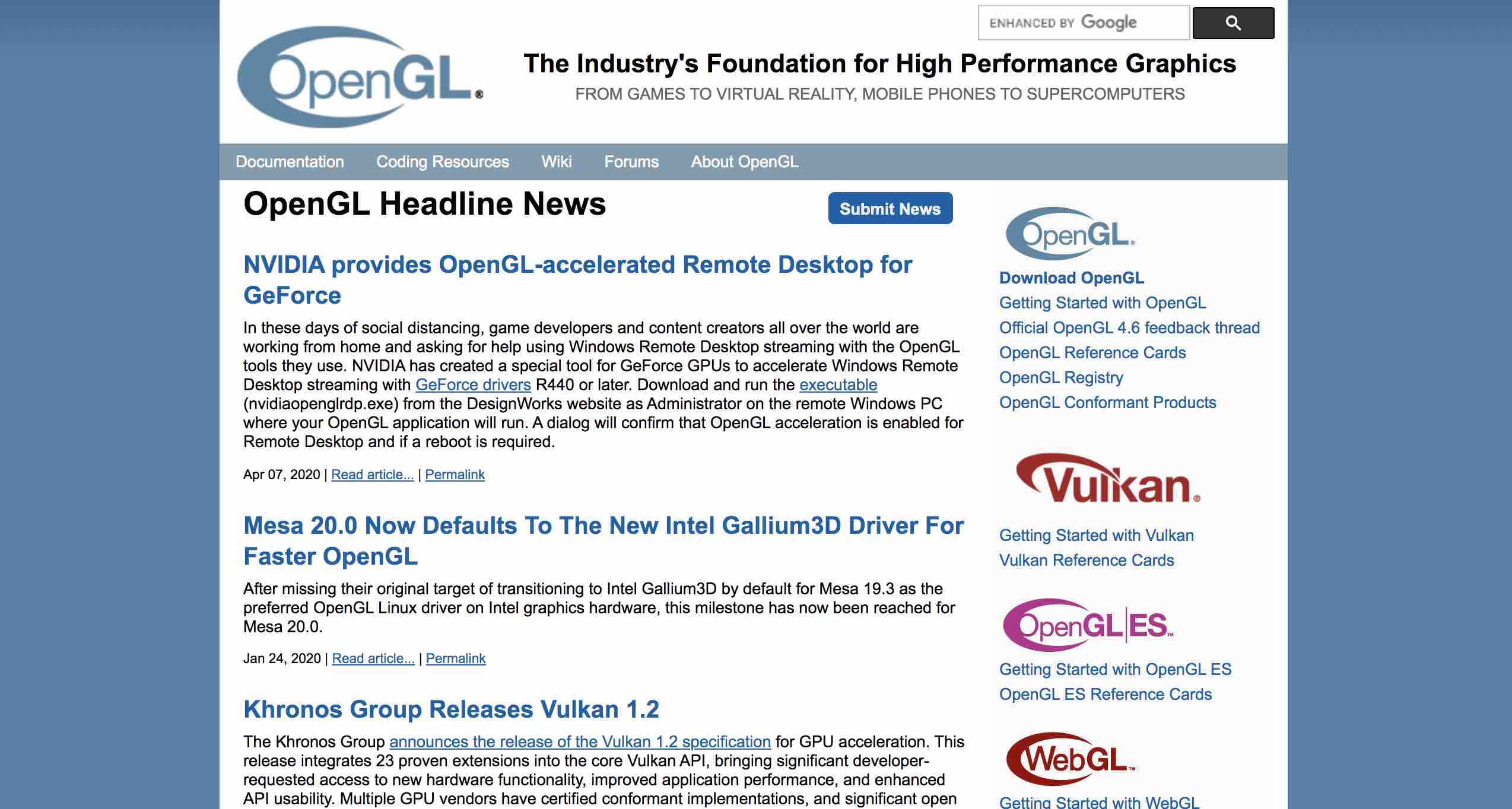}
        \end{minipage}%
    }%
    \centering
    \caption{Thumbnail images of Visualization-Focused Websites.  They are presented in the same order as Table~\ref{tab:websites}}
    \label{fig:websites}
\end{figure}

As part of our search for visualization resources, we examined web sites that focus on visualization and visual analytics.
\textbf{The Search Process:}
Our search process actually began while we compiled the list of Information Visualization books~\cite{Rees2019a}.
During this multi-year search process, we made note of authors with high-quality,
accompanying web pages that featured a substantial collection of visualization resources (and not simply a web page for marketing their own work).
We also searched for ``Data Visualization web sites" and ``Visual Analytics web sites".
We did not include all websites that we found related to these topics.
Each web site featured in this section went through a quality and appropriateness checklist in order to be included.
\textbf{Inclusion Criteria:}
One of the criteria we used is that each web site features a substantial and well organized collection of visualization resources that offers value to the reader audience and is not simply for commercial purposes.
Another criteria we included was that the web site content is up to date, updated within the last year.
We checked for quality web site organization with clear content categories as well as offering a substantive amount of valuable visualization resources.
These resources include:
\begin{enumerate}
    \item Guidance on choosing a visual design
    \item Blog(s)
    \item Visualization training and Educational Resources
    \item Events such as conferences, workshops etc
    \item Visualization Related Publications
    \item Visualization Book Collections and Recommendations
    \item Visualization Tools and Software
    \item Data Sources
    \item Case Studies and Examples
    \item Links to Related Web Pages
\end{enumerate}
The list of web sites we found that meet our quality criteria are summarized in Tables~\ref{tab:websites} and~\ref{tab:websites_resources}.
In general, we do not include web pages of individual authors simply listing their own publications.
Generally, we do not include solely for-profit news articles.
An overview of the websites is provided by Figure~\ref{fig:websites}, and Tables~\ref{tab:websites} and~\ref{tab:websites_resources}.

\begin{table*}[!tb]
    \caption{Website References: A summary of the SurVis web pages described in Section~\ref{sec:websites}. These web sites feature a quality collection of visualization-related resources. See Figure~\ref{fig:websites} for images of the websites}

    \centering

    \resizebox{\textwidth}{!}{
        \begin{tabulary}{\textwidth}{|l|l|}
            \hline
            \textbf{Web Page Name}  &  \textbf{URL}  \\ \hline
            \textbf{Depict Data Studio} \cite{AnnK.Emery} & \url{https://depictdatastudio.com/charts/}   \\ \hline
            \textbf{DVS}\cite{DataVisualizationSocietya} & \url{https://www.datavisualizationsociety.com/}   \\ \hline
            \textbf{DVP}\cite{Ferdio} & \url{https://datavizproject.com/}  \\ \hline
            \textbf{From Data to Viz}\cite{YanHoltz} & \url{https://www.data-to-viz.com/}  \\ \hline
            \textbf{InformationIsBeautiful.net}\cite{DavidMcCandless} & \url{https://informationisbeautiful.net/}   \\ \hline
            \textbf{The InfoVis Wiki page} \cite{InfoVis:Wikiteam} & \url{https://infovis-wiki.net/wiki/Main_Page}   \\ \hline
            \textbf{SankeyMATIC} \cite{SteveBogar} & \url{http://sankeymatic.com/build/}  \\ \hline
            \textbf{Seeing Data} \cite{SeeingDataprojectteam}& \url{http://seeingdata.org/}  \\ \hline
            \textbf{Tableau Public} \cite{Tableaub} & \url{https://public.tableau.com/en-us/gallery/?tab=viz-of-the-day&type=viz-of-the-day}  \\ \hline
            \textbf{Data Visualisation Catalogue} \cite{SeverinoRibecca} & \url{https://datavizcatalogue.com/}  \\ \hline
            \textbf{Visual Vocabulary}\cite{Kriebel} & \url{https://ft-interactive.github.io/visual-vocabulary/} \\ \hline
            \textbf{Visualisingdata.com} \cite{AndyKirk} & \url{https://www.visualisingdata.com/}  \\ \hline
            \textbf{VRVis Conference Calendar} \cite{HelwigHause}& \url{https://confcal.vrvis.at/conferences/} \\ \hline
            \textbf{VizHub} \cite{DatavisTech} & \url{https://vizhub.com/}  \\ \hline
            \textbf{Visualizing.org} \cite{GE}& \url{https://www.visualizing.org/} \\ \hline
            \textbf{OpenGL}\cite{KhronosGroup} & \url{https://www.opengl.org/}  \\ \hline
            \textbf{The Visualization Universe}\cite{visuniverse} & \url{http://visualizationuniverse.com/}  \\ \hline

        \end{tabulary}
    }
    \label{tab:websites}
\end{table*}

\begin{table*}[!tb]
    \caption{Website References: A summary of the web pages and relevant resources described in Section~\ref{sec:websites}.}
    \centering
    \resizebox{\textwidth}{!}{
        \begin{tabulary}{\textwidth}{|l|*{10}{c|}}
            \hline
            \textbf{ }  & \textbf{Design Guidance}
            & \textbf{Blog(s) }
            & \textbf{Training}
            & \textbf{Events}
            & \textbf{Related Publications }
            & \textbf{Book Collections}
            & \textbf{Tools and Software}
            & \textbf{Data Sources}
            & \textbf{Case Studies}
            & \textbf{Related Web Pages}  \\ \hline\hline

            \textbf{Depict Data Studio} \cite{AnnK.Emery}  &  $\surd$ & $\surd$ & $\surd$ & $\surd$ &   & $\surd$  & \textbf{}  & \textbf{}  & \textbf{}  & \textbf{}    \\ \hline

            \textbf{DVS}\cite{DataVisualizationSocietya}  & $\surd$ & $\surd$ &   $\surd$ & $\surd$ &  $\surd$ & $\surd$  & \textbf{}  &  $\surd$  &  $\surd$  &  $\surd$  \\ \hline

            \textbf{DVP}\cite{Ferdio}  &  $\surd$ & \textbf{} &  \textbf{}&  \textbf{}&  \textbf{} & \textbf{}  & $\surd$ & \textbf{}  & \textbf{}  & \textbf{}   \\ \hline

            \textbf{From Data to Viz}\cite{YanHoltz}  & $\surd$  & \textbf{} &  \textbf{}&  \textbf{}&  \textbf{} & \textbf{}  &  $\surd$  & $\surd$ & \textbf{}  & \textbf{}   \\ \hline

            \textbf{Information Is Beautiful.net}\cite{DavidMcCandless} &  $\surd$ &  $\surd$ & $\surd$ &  \textbf{}&  \textbf{} & \textbf{}  & $\surd$ & \textbf{}  & $\surd$ & \textbf{} \\ \hline

            \textbf{The InfoVis Wiki page} \cite{InfoVis:Wikiteam} &   \textbf{} & \textbf{} & $\surd$ &  $\surd$ &  $\surd$ & \textbf{}  & \textbf{}  & \textbf{}  & \textbf{}  & $\surd$ \\ \hline

            \textbf{SankeyMATIC} \cite{SteveBogar}  &  $\surd$ & \textbf{} &  \textbf{}&  \textbf{}&  \textbf{} & \textbf{}  & \textbf{} & $\surd$ & \textbf{}  & \textbf{}   \\ \hline

            \textbf{Seeing Data} \cite{SeeingDataprojectteam} & $\surd$ & $\surd$ & $\surd$ & $\surd$ & \textbf{} & \textbf{}  & \textbf{}  & \textbf{}  & \textbf{}  & \textbf{}   \\ \hline

            \textbf{Tableau Public} \cite{Tableaub}  & $\surd$ & $\surd$ &  \textbf{}&  \textbf{}&  \textbf{} & \textbf{}  & $\surd$ & $\surd$ &  $\surd$  &  $\surd$ \\ \hline

            \textbf{Data Visualisation Catalogue} \cite{SeverinoRibecca}   &   $\surd$  &  $\surd$  &  $\surd$ & $\surd$ & $\surd$  &   & $\surd$  &  $\surd$  &  $\surd$ &  $\surd$   \\ \hline

            \textbf{Visual Vocabulary}\cite{Kriebel}  & $\surd$ & \textbf{} &  $\surd$ &  \textbf{}&  \textbf{} & \textbf{}  & \textbf{}  & \textbf{}  & \textbf{}  & \textbf{}   \\ \hline

            \textbf{Visualisingdata.com} \cite{AndyKirk}  & $\surd$ & $\surd$ &  $\surd$ &  \textbf{}&  \textbf{} &  & $\surd$ & \textbf{}  & $\surd$ & $\surd$ \\ \hline

            \textbf{VRVis Conference Calendar} \cite{HelwigHause} &  \textbf{} & \textbf{} &  \textbf{}& $\surd$ &  \textbf{} & \textbf{}  & \textbf{}  & \textbf{}  & \textbf{}  & \textbf{}  \\ \hline

            \textbf{VizHub} \cite{DatavisTech}  & $\surd$ & \textbf{} & $\surd$ &  \textbf{}&  \textbf{} & \textbf{}  & $\surd$ & \textbf{}  & $\surd$ & \textbf{}   \\ \hline

            \textbf{Visualizing.org} \cite{GE} & $\surd$ & $\surd$ &  & $\surd$ &  \textbf{} & \textbf{}  & $\surd$  &$\surd$  & $\surd$  & \\ \hline

            \textbf{OpenGL}\cite{KhronosGroup}  & $\surd$ & \textbf{} & $\surd$ &   $\surd$ & & & $\surd$ & $\surd$  & \textbf{}  &  $\surd$  \\ \hline

            \textbf{The Visualization Universe}\cite{visuniverse}
            & $\surd$           
            &                   
            &                   
            &                   
            &                   
            & $\surd$           
            & $\surd$           
            &                   
            &                   
            &                   
            \\ \hline
        \end{tabulary}
    }
    \label{tab:websites_resources}
\end{table*}

\textbf{Depict Data Studio} \cite{AnnK.Emery} hosts a website for presenting visualizations, reports, slideshows and dashboards to non-technical audiences. The website was created in 2014 in Florida. Content includes online courses, group training, live events information, and related blogs. The website also hosts an online tool called an interactive chart chooser providing guidance on the most suitable visualization types for a given set of data. The visual designs are categorized into 10 types: 1 point in time, 2 points in time, 3+ points in time, comparisons, correlation, distribution, exploratory, part-to-whole, and progress towards goals and relationships. Under each label, suitable visual layouts are provided with a brief introduction and several illustrations.

The data visualization society.com (\textbf{DVS}) web site \cite{DataVisualizationSocietya} is a platform for community building targeted at professionals and practitioners incorporating visual data analysis into their professional practice. Founded recently in 2019, and based primarily in North America, DVS is focused on building a community for sharing best practices, skills, and experiences. DVS content consists of video-based panels and tutorials, a survey of data visualization practitioners, a buddy program pairing more experienced practitioners with those less experienced, an online publication featuring visualization case studies, a collection of data sources, a schedule of meetups and events (including conferences and workshops), and a collection of awards related to data visualization. There are also data visualization challenges for those looking to practice their technical skills, a spreadsheet of visualization-related books, a list of visualization related courses ranging from a single day to a multi-year degree program, and a list of podcasts, blogs, and related web sites.

The Data Viz Project (\textbf{DVP}) \cite{Ferdio} is website created by a data visualization agency \cite{Ferdioa} in Copenhagen. The DVP is an online tool enabling users to find the most appropreate visual layouts for their data and learn how to create images. The website provides different perspectives to inspire users who struggle to creating their visual interface. Users can view visual designs according to the kind, input data, the functions to deliver the methods, as well as the shape of the visualization.

\textbf{From Data to Viz} \cite{YanHoltz} is an online chart selector for data visualization, created by Yan Holtz. The website serves as an online tool to explore and visualize data according to types of data and the sorts of visualization.

\textbf{InformationIsBeautiful.net} \cite{DavidMcCandless} is a website for data visualization techniques founded by David McCandless. The examples provided by the website are regularly updated and revised. The website provides a collection of visual designs on different topics: new, tech and digital, science and health, nature and climate, money and economy, ideas and perception, gender and diversity, food and drink, data visualization design, as well as arts and culture. Furthermore, there are also blogs, online training and recommendations on creating visualizations from scratch.

\textbf{The InfoVis Wiki page} \cite{InfoVis:Wikiteam} is an online community platform related to information visualization. The InfoVis project was created (and maintained) by the InfoVis Wiki team in 2003. The aim of the project is to provide up-to-date developments and information on the field of information visualization. The page adopts Wiki technology and is editable-by-anyone. The platform contributes news, a glossary, visualization design patterns, visualization events, web resources, publications, research and educational information to users. In addition, information about researchers, companies, jobs, techniques and software is provided.
Although this project does not generally include Wikipedia pages (due to quantity), we make an exception for the InfoVis Wiki page due to its substantial collection of resources.

The \textbf{SankeyMATIC} \cite{SteveBogar} is an open source javascript web application produced by Steve Bogart. It provides a GUI to the D3.js Sankey library and enables users to create a Sankey diagram online.

\textbf{Seeing Data} \cite{SeeingDataprojectteam} is a website hosting a collection of data visualization projects. The online resource was founded by the Seeing Data project team University of Sheffield. The aim of this project is to help users understand how to explore data in the form of images. The website provides guidance in developing visualization literacy, relating to data through visualization, and innovative news on data visualization and blogs.

\textbf{Tableau Public} \cite{Tableaub} is an online platform for sharing and exploring data visualizations. It is a free product owned by Tableau Software, LLC. The web site hosts a collection of visualization resources created by Tableau Public tools, and a list of authors working on visualization. The web site also contributes free visualization tools for users.

The \textbf{Data Visualisation Catalogue} \cite{SeverinoRibecca} is a project developed by Severino Ribecca. The project aims to create a library that provides a variety of information visualization resources. The website hosts a collection of data sources, dataviz blogs, code-based chart generation tools, chart generation WebApps, and examples showcasing visual samples and infographics, libraries, publications, podcasts, conferences, etc.

\textbf{Visual Vocabulary} \cite{Kriebel} is an online platform created by Andy Kriebel. The webpage is also featured by Tableau Public \cite{Tableaua}. It serves as a starting point for making informative and meaningful data visualisations. Visual Vocabulary provides different strategies to explore data: deviation, correlation, ranking, distribution, change over time, part-to whole, magnitude, spatial and flow.

\textbf{Visualisingdata.com} \cite{AndyKirk} is a website featuring a collection of tools and resources related to data visualization. The website was founded by Andy Kirk in 2010 based in Leeds. Apart from podcasts, blogs, training courses and books, the website also hosts a collection of visualization resources. The visualizations in the collection contains data handling, applications, programming, web-based tools, qualitative, mapping, specialist and color mapping.

The \textbf{VRVis Conference Calendar} \cite{HelwigHause} is a website for searching data visualization conferences. The website was founded on the basis of Helwig’s Conference Calendar and created by Helwig Hauser at the Vienna University of Technology in 1997. The website hosts a collection of more than 1000 data visualization events since 1972.

The \textbf{VizHub} \cite{DatavisTech} is a web site created by the Datavis Tech Inc providing courses and open source code examples for data visualization. VisHub also serves as an online tool to create and develop visualizations using web technologies such as HTML, CSS, JavaScript and JSX(React). In addition, the web site can be used as an educational platform to produce an online course in data visualization.

\textbf{Visualizing.org} \cite{GE} is an online community created by GE and the Seed Media Group. The website focuses on visualizing important events for designers, organizations, teachers and schools, as well as audiences who are trying to exploring data visualization and infographics.

\textbf{OpenGL} \cite{KhronosGroup} is a graphics standard for developing portable, interactive 2D and 3D graphics applicatons. It was created by the Khronos Group Inc in 1992. In the website, books, tutorials, online coding examples, and coding seminars, are provided.

\textbf{The Visualization Universe} \cite{visuniverse} is a collaboration project between Google News Lab \cite{GoogleNewsInitiative} and Adioma\cite{Adioma}. The project is an analytical, up-to-date visualization resource that aims to present the state of the data visualization based on the current user's search and interest activities. The website features three content categories: tools, books, and charts.

\section{Conclusion}
In this paper, we present a novel collection of visualization resources. We classify the resources according to type. We compile and provide the descriptions, examples and information about the curators of each resource collection. This paper serves as a novel starting point and comprehensive overview on visualization resources for students, researchers, and practitioners with an interest in data visualization and visual analytics.

There are many potential future work directions which can be explored. For this paper, we only focus on collections of resources for literature and websites. However, there are many other types of resources such as software, online videos, and other tools for programmers. Therefore, we would like to expand our survey with more categories. We also think that collections of resources that focus on special topics of data visualization will make the survey more comprehensive.

\section{Acknowledgements}
This project was support in part by 
the Engineering and Physical Sciences Resource Council (EPSRC EP/S010238/1).

\bibliographystyle{plainurl}
\bibliography{reference}

\begin{thebibliography}{10}

\bibitem{Adioma}
{Adioma - Infographic Maker With Timelines, Grids and Icons}.
\newblock URL: \url{https://adioma.com/}.

\bibitem{GoogleNewsInitiative}
{Google News Initiative Training Center}.
\newblock URL: \url{https://newsinitiative.withgoogle.com/training/}.

\bibitem{visuniverse}
{ Anna Vital, Mark Vital}.
\newblock {The Visualization Universe}.
\newblock URL: \url{http://visualizationuniverse.com//}.

\bibitem{Aigner2011}
Wolfgang Aigner, Silvia Miksch, Heidrun Schumann, and Christian Tominski.
\newblock {Survey of Visualization Techniques}.
\newblock In {\em Visualization of Time-Oriented Data}, pages 147--254.
  Springer, London, 2011.
\newblock URL:
  \url{http://vcg.informatik.uni-rostock.de/~ct/timeviz/timeviz.html}.

\bibitem{Aigner2011a}
Wolfgang Aigner, Silvia Miksch, Heidrun Schumann, and Christian Tominski.
\newblock {\em {Visualization of Time-Oriented Data}}.
\newblock 2011.
\newblock URL: \url{https://www.timeviz.net/}.

\bibitem{AlbertoCairo}
{Alberto Cairo}.
\newblock {The Functional Art: An Introduction to Information Graphics and
  Visualization: Download the Datasaurus: Never trust summary statistics alone;
  always visualize your data}.
\newblock URL: \url{http://www.thefunctionalart.com/}.

\bibitem{Alharbi2018}
Mohammad Alharbi and Robert~S. Laramee.
\newblock {SoS TextVis: A survey of surveys on text visualization}.
\newblock {\em Computer Graphics and Visual Computing, CGVC 2018}, pages
  143--152, 2018.

\bibitem{Alharbi2019a}
Mohammad Alharbi and Robert~S. Laramee.
\newblock {SoS textvis: An extended survey of surveys on text visualization}.
\newblock {\em Computers}, 8(1):1--18, 2019.

\bibitem{Alharbi2017}
Naif Alharbi, Mohammad Alharbi, Xavier Martinez, Michael Krone, Alexander~S.
  Rose, Marc Baaden, Robert~S. Laramee, and Matthieu Chavent.
\newblock {Molecular Visualization of Computational Biology Data: A Survey of
  Surveys}.
\newblock {\em Eurographics Conference on Visualization (EuroVis) 2017}, pages
  2--6, 2017.
\newblock URL:
  \url{http://www.cs.swan.ac.uk/~csbob/research/star/mdv/sos/alharbi17parallelCoords.html}.

\bibitem{AndyKirk}
{Andy Kirk}.
\newblock {Visualising Data}.
\newblock URL: \url{https://www.visualisingdata.com/}.

\bibitem{AnnK.Emery}
{Ann K. Emery}.
\newblock {Interactive Chart Chooser | Depict Data Studio}.
\newblock URL: \url{https://depictdatastudio.com/charts/}.

\bibitem{Anscombe1973}
F.~J. Anscombe.
\newblock {Graphs in Statistical Analysis}.
\newblock {\em The American Statistician}, 27(1):17, feb 1973.

\bibitem{Beck2014}
Fabian Beck, Michael Burch, Stephan Diehl, and Daniel Weiskopf.
\newblock {The State of the Art in Visualizing Dynamic Graphs}.
\newblock {\em Proceedings State of the Art Reports (STARs)}, pages 83--103,
  2014.
\newblock URL: \url{http://dynamicgraphs.fbeck.com/}.

\bibitem{Beck2016b}
Fabian Beck, Sebastian Koch, and Daniel Weiskopf.
\newblock {Visual Analysis and Dissemination of Scientific Literature
  Collections with SurVis}.
\newblock {\em IEEE Transactions on Visualization and Computer Graphics},
  22(1):180--189, jan 2016.

\bibitem{Beck2017a}
Fabian Beck and Daniel Weiskopf.
\newblock {Word-sized graphics for scientific texts}.
\newblock {\em IEEE Transactions on Visualization and Computer Graphics},
  23(6):1576--1587, 2017.
\newblock URL: \url{http://sparklines-literature.fbeck.com/}.

\bibitem{Borgo}
Rita Borgo, Luana Micallef, Benjamin Bach, Fintan McGee, and Bongshin Lee.
\newblock {Information Visualization Evaluation Using Crowdsourcing}, 2018.
\newblock URL: \url{https://crowdsourcing4vis.github.io/}.

\bibitem{GoodReads}
Otis Chandler and Elizabeth~Khuri Chandler.
\newblock {Information Visualization Books, GoodReads.com}.
\newblock URL:
  \url{https://www.goodreads.com/shelf/show/information-visualization}.

\bibitem{Chatzimparmpas2020}
A.~Chatzimparmpas, R.~M. Martins, I.~Jusufi, K.~Kucher, F.~Rossi, A.~Kerren,
  R.~M. Martins, I.~Jusufi, K.~Kucher, F.~Rossi, and A.~Kerren.
\newblock {The State of the Art in Enhancing Trust in Machine Learning Models
  with the Use of Visualizations}.
\newblock {\em Eurovis 2020}, 39(3):713--756, jun 2020.
\newblock URL: \url{https://trustmlvis.lnu.se/}.

\bibitem{chen2020vis30k}
Jian Chen, Meng Ling, Rui Li, Petra Isenberg, Tobias Isenberg, Michael
  Sedlmair, Torsten M{\"o}ller, Robert~S Laramee, Han-Wei Shen, Katharina
  W{\"u}nsche, et~al.
\newblock Vis30k: A collection of figures and tables from $\{$IEEE$\}$
  visualization conference publications.
\newblock {\em arXiv preprint arXiv:2101.01036}, 2020.
\newblock URL: \url{https://arxiv.org/abs/2101.01036}.

\bibitem{DataVisualizationSocietya}
{Data Visualization Society}.
\newblock {Data Visualization Society}.
\newblock URL: \url{https://www.datavisualizationsociety.com/}.

\bibitem{Society}
{Data Visualization Society} and {Information is Beautiful}.
\newblock {Data-Visualization Books Everyone Should Read - Extended List}.
\newblock URL: \url{https://airtable.com/shrugbQMDGVNvArMT/tblSrU1fNAykSMyXU}.

\bibitem{DataVisualizationSociety}
{Data Visualization Society} and {Information is Beautiful}.
\newblock {Data-Visualization Books Everyone Should Read — Information is
  Beautiful}.
\newblock URL:
  \url{https://informationisbeautiful.net/visualizations/dataviz-books/}.

\bibitem{DatavisTech}
{Datavis Tech}.
\newblock {VizHub}.
\newblock URL: \url{https://vizhub.com/}.

\bibitem{DavidMcCandless}
{David McCandless}.
\newblock {Information is Beautiful}.
\newblock URL: \url{https://informationisbeautiful.net/}.

\bibitem{Dumas2014}
Maxime Dumas, Michael~J McGuffin, and Victoria~L Lemieux.
\newblock {Financevis.net : A Visual Survey of Financial Data Visualizations}.
\newblock In {\em Poster Abstracts of IEEE conference on visualization},
  volume~2, 2014.
\newblock URL: \url{http://financevis.net/}.

\bibitem{Federico2016c}
Paolo Federico, Florian Heimerl, Steffen Koch, and Silvia Miksch.
\newblock {A Survey on Visual Approaches for Analyzing Scientific Literature
  and Patents}.
\newblock {\em IEEE Transactions on Visualization and Computer Graphics},
  PP(99):2179--2198, 2016.
\newblock URL: \url{http://ieg.ifs.tuwien.ac.at/~federico/LiPatVis/}.

\bibitem{Tableaua}
FT.
\newblock {Visual Vocabulary | Tableau Public}, 2018.
\newblock URL: \url{https://public.tableau.com/en-us/gallery/visual-vocabulary
  https://public.tableau.com/en-us/s/gallery/visual-vocabulary}.

\bibitem{GE}
GE and {Seed Media Group}.
\newblock {Visualizing.org}.
\newblock URL: \url{https://www.visualizing.org/}.

\bibitem{HelwigHause}
{Helwig Hauser}.
\newblock {VRVis Conference Calendar, Vrvis GmbH}.
\newblock URL: \url{https://confcal.vrvis.at/conferences/}.

\bibitem{InfoVis:Wikiteam}
{InfoVis: Wiki team}.
\newblock {InfoVis:Wiki}.
\newblock URL: \url{https://infovis-wiki.net/wiki/Main_Page}.

\bibitem{Isaacs2014}
Katherine~E Isaacs, Alfredo Gim{\'{e}}nez, Ilir Jusufi, Todd Gamblin, Abhinav
  Bhatele, Martin Schulz, Bernd Hamann, and Peer-timo Bremer.
\newblock {State of the Art of Performance Visualization}.
\newblock {\em EuroVis 2014}, pages 141--160, 2014.
\newblock URL: \url{http://hdc.cs.arizona.edu/people/kisaacs/STAR/}.

\bibitem{Isenberga}
Isenberg.
\newblock {VIS Contribution Types | List of contribution types in visualization
  research along with exemplar papers}.
\newblock URL: \url{https://vis-contribution-types.github.io/}.

\bibitem{Isenberg2017a}
Petra Isenberg, Florian Heimerl, Steffen Koch, Tobias Isenberg, Panpan Xu,
  Charles~D. Stolper, Michael Sedlmair, Jian Chen, Torsten Moller, and John
  Stasko.
\newblock {Vispubdata.org: A Metadata Collection About IEEE Visualization (VIS)
  Publications}.
\newblock {\em IEEE Transactions on Visualization and Computer Graphics},
  23(9):2199--2206, sep 2017.
\newblock URL: \url{https://sites.google.com/site/vispubdata/home}.

\bibitem{Isenberg2017b}
Petra Isenberg, Tobias Isenberg, Michael Sedlmair, Jian Chen, and Torsten
  Moller.
\newblock {Visualization as Seen through its Research Paper Keywords}.
\newblock {\em IEEE Transactions on Visualization and Computer Graphics},
  23(1):771--780, jan 2017.
\newblock URL: \url{http://keyvis.org/}.

\bibitem{isenberg2013systematic}
Tobias Isenberg, Petra Isenberg, Jian Chen, Michael Sedlmair, and Torsten
  M{\"o}ller.
\newblock A systematic review on the practice of evaluating visualization.
\newblock {\em IEEE Transactions on Visualization and Computer Graphics},
  19(12):2818--2827, 2013.
\newblock URL: \url{10.1109/TVCG.2013.126}.

\bibitem{Jena2020}
Amit Jena, Ulrich Engelke, Tim Dwyer, Venkatesh Raiamanickam, and Cecile Paris.
\newblock {Uncertainty Visualisation: An Interactive Visual Survey}.
\newblock {\em IEEE Pacific Visualization Symposium}, 2020-June:201--205, 2020.
\newblock URL: \url{https://amitjenaiitbm.github.io/uncertaintyVizBrowser/}.

\bibitem{Kehrer2013a}
Johannes Kehrer and Helwig Hauser.
\newblock {Visualization and Visual Analysis of Multifaceted Scientific Data: A
  Survey}.
\newblock {\em IEEE Transactions on Visualization and Computer Graphics},
  19(3):495--513, mar 2013.
\newblock URL: \url{https://multivis.net/}.

\bibitem{Kerren2017a}
Andreas Kerren, Kostiantyn Kucher, Yuan-Fang Li, and Falk Schreiber.
\newblock {BioVis Explorer: A visual guide for biological data visualization
  techniques}.
\newblock {\em PLOS ONE}, 12(11):e0187341, nov 2017.
\newblock URL: \url{https://biovis.lnu.se/}.

\bibitem{KhronosGroup}
{Khronos Group}.
\newblock {OpenGL - The Industry Standard for High Performance Graphics}.
\newblock URL: \url{https://www.opengl.org/}.

\bibitem{Kriebel}
Andy Kriebel.
\newblock {Visual Vocabulary, Github.io}.
\newblock URL: \url{https://ft-interactive.github.io/visual-vocabulary/}.

\bibitem{Kucher2015}
Kostiantyn Kucher and Andreas Kerren.
\newblock {Text visualization techniques: Taxonomy, visual survey, and
  community insights}.
\newblock In {\em IEEE Pacific Visualization Symposium}, volume 2015-July,
  pages 117--121. IEEE, apr 2015.
\newblock URL: \url{https://textvis.lnu.se/}.

\bibitem{Kucher2018a}
Kostiantyn Kucher, Carita Paradis, and Andreas Kerren.
\newblock {The state of the art in sentiment visualization}.
\newblock {\em Computer Graphics Forum}, 37(1):71--96, feb 2018.
\newblock URL: \url{https://sentimentvis.lnu.se/}.

\bibitem{Lam2012a}
Heidi Lam, Enrico Bertini, Petra Isenberg, Catherine Plaisant, and
  S.~Carpendale.
\newblock {Empirical Studies in Information Visualization: Seven Scenarios}.
\newblock {\em IEEE Transactions on Visualization and Computer Graphics},
  18(9):1520--1536, sep 2012.
\newblock URL:
  \url{https://docs.google.com/document/d/1uSius4qLHHAERhUtuMrfJ5nMyE-X9cot3ewUdZfR0lQ/edit?authkey=CIjs31Q}.

\bibitem{ling2020deeppapercomposer}
Meng Ling and Jian Chen.
\newblock {DeepPaperComposer}: A simple solution for training data preparation
  for parsing research papers.
\newblock In {\em Proc.\ EMNLP/Scholarly Document Processing}, pages 91--96,
  Stroudsburg, PA, USA, 2020. ACL.
\newblock URL: \url{10.18653/v1/2020.sdp-1.10}.

\bibitem{Liu2017}
Shusen Liu, Dan Maljovec, Bei Wang, Peer~Timo Bremer, and Valerio Pascucci.
\newblock {Visualizing High-Dimensional Data: Advances in the Past Decade}.
\newblock {\em IEEE Transactions on Visualization and Computer Graphics},
  23(3):1249--1268, 2017.
\newblock URL: \url{http://www.sci.utah.edu/~shusenl/highDimSurvey/website/}.

\bibitem{Lu2017a}
Yafeng Lu, Rolando Garcia, Brett Hansen, Michael Gleicher, and Ross
  Maciejewski.
\newblock {The State-of-the-Art in Predictive Visual Analytics}.
\newblock {\em Computer Graphics Forum}, 36(3):539--562, 2017.
\newblock URL: \url{http://104.196.253.120/pva_browser/}.

\bibitem{Matejka2017}
Justin Matejka and George Fitzmaurice.
\newblock {Same Stats, Different Graphs, Autodesk.com}, 2017.
\newblock URL: \url{https://www.autodesk.com/}.

\bibitem{Justin}
Justin Matejka and George Fitzmaurice.
\newblock Same stats, different graphs: Generating datasets with varied
  appearance and identical statistics through simulated annealing.
\newblock In {\em Proceedings of the 2017 CHI Conference on Human Factors in
  Computing Systems}, CHI '17, page 1290–1294, New York, NY, USA, 2017. ACM.
\newblock URL: \url{https://doi.org/10.1145/3025453.3025912}.

\bibitem{McNabb2017c}
Liam McNabb and Robert~S. Laramee.
\newblock {Survey of Surveys (SoS) ‐ Mapping The Landscape of Survey Papers
  in Information Visualization}.
\newblock {\em Computer Graphics Forum}, 36(3):589--617, jun 2017.
\newblock URL: \url{http://sos.swansea.ac.uk/}.

\bibitem{Min}
Chen Min and {Min Chen}.
\newblock {CGF State of the Art Reports - Min Chen, University of Oxford}.
\newblock URL:
  \url{https://sites.google.com/site/drminchen/cgf-info/cgf-stars}.

\bibitem{Ferdioa}
Birger Morgenstjerne, Jeppe Morgenstjerne, Tobias Jeppesen, Klaudia Gal,
  Nicolas Maravitti, Nelly Boyanova, Chris Ntantos, and Beatriz Rocha.
\newblock ferdio.
\newblock URL: \url{https://www.ferdio.com/}.

\bibitem{Ferdio}
Birger Morgenstjerne, Jeppe Morgenstjerne, Tobias Jeppesen, Klaudia Gal,
  Nicolas Maravitti, Nelly Boyanova, Chris Ntantos, and Beatriz Rocha.
\newblock {Data Viz Project | Collection of data visualizations to get inspired
  and finding the right type.}, 2020.
\newblock URL: \url{https://datavizproject.com/}.

\bibitem{Nobre2019b}
C.~Nobre, M.~Meyer, M.~Streit, and A.~Lex.
\newblock {The state of the art in visualizing multivariate networks}.
\newblock {\em Computer Graphics Forum}, 38(3):807--832, 2019.
\newblock URL: \url{https://vdl.sci.utah.edu/mvnv/}.

\bibitem{Nusrat2016a}
Sabrina Nusrat and Stephen Kobourov.
\newblock {The State of the Art in Cartograms}.
\newblock {\em Computer Graphics Forum}, 35(3):619--642, 2016.

\bibitem{SeeingDataprojectteam}
Seeing~Data Project.
\newblock {Seeing Data Summary and Objectives, Seeingdata.org}.
\newblock URL: \url{http://seeingdata.org/}.

\bibitem{Rees2019a}
D.~Rees and R.~S. Laramee.
\newblock {A Survey of Information Visualization Books}.
\newblock {\em Computer Graphics Forum}, 38(1):610--646, 2019.
\newblock URL: \url{http://visbooks.swansea.ac.uk/}.

\bibitem{SeverinoRibecca}
Severino Ribecca.
\newblock {The Data Visualisation Catalogue}.
\newblock pages 1--4, 2015.
\newblock URL: \url{https://datavizcatalogue.com/}.

\bibitem{Sandra2019}
Durcevi Sandra.
\newblock {The 18 Best Data Visualization Books You Should Read, Datapine.com},
  dec 2019.
\newblock URL:
  \url{https://www.datapine.com/blog/best-data-visualization-books/}.

\bibitem{Schulz2011}
Hans-Jorg Schulz.
\newblock {Treevis.net: A Tree Visualization Reference}.
\newblock {\em IEEE Computer Graphics and Applications}, 31(6):11--15, nov
  2011.
\newblock URL: \url{https://treevis.net/}.

\bibitem{FabioSouto}
Fabio Souto.
\newblock {Awesome Dataviz}.
\newblock URL: \url{https://github.com/fasouto/awesome-dataviz}.

\bibitem{Stasko2013}
John Stasko, Jaegul Choo, Yi~Han, and Mengdie Hu.
\newblock {Citevis: Exploring conference paper citation data visually}.
\newblock {\em Posters of IEEE InfoVis.}, pages 2--3, 2013.
\newblock URL: \url{https://www.cc.gatech.edu/gvu/ii/citevis/}.

\bibitem{SteveBogar}
Bogart Steve.
\newblock {SankeyMATIC (BETA): Build a diagram}, 2017.
\newblock URL: \url{http://sankeymatic.com/build/}.

\bibitem{Tableau}
Tableau.
\newblock {12 great books about data visualisation | Tableau Software}.
\newblock URL:
  \url{https://www.tableau.com/en-gb/learn/articles/books-about-data-visualization}.

\bibitem{Tableaub}
Tableau.
\newblock {Galery | Tableau Public}, 2020.
\newblock URL:
  \url{https://public.tableau.com/en-us/gallery/?tab=viz-of-the-day&type=viz-of-the-day}.

\bibitem{TimothyKing}
{Timothy King}.
\newblock {The Top 30 Best Data Visualization Books on Our Reading List}.
\newblock URL:
  \url{https://solutionsreview.com/business-intelligence/best-data-visualization-books-on-amazon-you-should-read/}.

\bibitem{UsabiliTEST}
UsabiliTEST.
\newblock {Usability testing: Card Sorting, Prioritization Matrix {\&} SUS.}
\newblock URL: \url{https://www.usabilitest.com/}.

\bibitem{Vehlow2015}
Corinna Vehlow, Fabian Beck, and Daniel Weiskopf.
\newblock {The state of the art in visualizing group structures in graphs}.
\newblock {\em Eurovis: Eurographics/IEEE Symposium on Visualization},
  2015:21--40, 2015.
\newblock URL: \url{http://cartogram.cs.arizona.edu/survis-cartogram/}.

\bibitem{Windhager2019a}
Florian Windhager, Paolo Federico, Gunther Schreder, Katrin Glinka, Marian
  Dork, Silvia Miksch, and Eva Mayr.
\newblock {Visualization of Cultural Heritage Collection Data: State of the Art
  and Future Challenges}.
\newblock {\em IEEE Transactions on Visualization and Computer Graphics},
  25(6):2311--2330, 2019.
\newblock URL: \url{http://collectionvis.org/}.

\bibitem{YanHoltz}
{Yan Holtz}.
\newblock {From data to Viz | Find the graphic you need}.
\newblock URL: \url{https://www.data-to-viz.com/}.

\end{thebibliography}

\end{document}